\documentclass[11pt]{article}
\usepackage{jheppub}
\usepackage{graphicx}
\usepackage{lipsum}
\usepackage{slashed}
\usepackage{mathtools}
\usepackage{bm}
\usepackage[all]{hypcap}
\usepackage{tikz}
\usepackage{tikz-feynman}
\usepackage{tikz-cd, quiver}
\usepackage{bbm}
\usepackage{xcolor}
\usepackage{subcaption}
\usepackage{chngpage}
\usepackage[font=small,labelfont=bf]{caption}

\usepackage{amsthm}

\theoremstyle{definition}

\def\Z{\mathbb{Z}}

\graphicspath{ {./Figs/} }
\usepackage{xspace}

\definecolor{myred}{RGB}{255,90,90}
\definecolor{mygreen}{rgb}{0.0,0.55,0.45}
\usepackage{amscd,amsxtra,mathrsfs,graphics,graphicx,amsthm,epsfig,bm,longtable,float,color,tikz,mathtools,xfrac,footnote,rotating,lscape,wrapfig}
\usepackage{simpler-wick}
\usepackage[makeroom]{cancel}
\usepackage[utf8]{inputenc}
\usepackage{bm}
\usepackage{physics}
\usepackage{caption,subcaption}
\usetikzlibrary{arrows}
\usepackage{empheq}
\usepackage{array}   
\newcolumntype{C}{>{$}c<{$}} 

\newcommand{\be}{\begin{equation}}
\newcommand{\ee}{\end{equation}}

\newcommand{\ul}[1]{\underline{#1}}

\def\so{\mathfrak{so}}

\def\su{\mathfrak{su}}
\def\SUL{\text{SU}(2)_L}

\def\U{\text{U}(1)}

\def\sp{\mathfrak{sp}}

\def\g{\mathfrak{g}}
\def\a{\mathfrak{a}}
\def\b{\mathfrak{b}}
\def\sm{\mathfrak{sm}}
\def\u{\mathfrak{u}(1)}
\def\h{\mathfrak{h}}
\def\so{\mathfrak{so}}
\def\su{\mathfrak{su}}
\def\sp{\mathfrak{sp}}
\def\hor{\mathfrak{hor}}
\def\ZP{Z^\prime}
\def\ua{\underline{a}}
\def\Qmax{Q_{\text{max}}}

\def\Tr{\text{Tr}}

\def\Q{\mathbb{Q}}
\def\C{\mathbb{C}}
\def\S{\mathbb{S}}

\begin{document}

\title{Flatland: 
abelian extensions of the Standard Model with semi-simple completions}
\date{}
\author[a]{Joe Davighi}
\author[b]{and Joseph Tooby-Smith}

\affiliation[a]{Physik-Institut, Universit\"at Z\"urich, CH 8057 Z\"urich, Switzerland}
\emailAdd{joe.davighi@physik.uzh.ch}
\affiliation[b]{Department of Physics, LEPP, Cornell University, Ithaca, NY 14853, USA}
\emailAdd{j.tooby-smith@cornell.edu}

\abstract{We parametrise the space of all possible flavour non-universal $\mathfrak{u}(1)_X$ extensions of the Standard Model that embed inside anomaly-free semi-simple gauge theories, including up to three right-handed neutrinos. More generally, we parametrise all abelian extensions ({\em i.e.} by any number of $\mathfrak{u}(1)$'s) of the SM with such semi-simple completions. The resulting space of abelian extensions is a collection of planes of dimensions $\leq 6$.
Numerically, we find that roughly $2.5\%$ of anomaly-free $\mathfrak{u}(1)_X$ extensions of the SM with a maximum charge ratio of $\pm 10$ can be embedded in such semi-simple gauge theories. Any vector-like anomaly-free abelian extension embeds (at least) inside $\g = \mathfrak{su}(12)\oplus \mathfrak{su}(2)_L\oplus \mathfrak{su}(2)_R$.
We also provide a simple computer program that tests whether a given $\mathfrak{u}(1)_{X^1}\oplus \mathfrak{u}(1)_{X^2}\oplus \dots$ charge assignment has a semi-simple completion and, if it does, outputs a set of maximal gauge algebras in which the $\sm\oplus\mathfrak{u}(1)_{X^1}\oplus \mathfrak{u}(1)_{X^2}\oplus \dots$ model may be embedded. We hope this is a useful tool in pointing the way from $\sm \oplus\mathfrak{u}(1)_{X^1}\oplus \mathfrak{u}(1)_{X^2}\oplus \dots$ models, which have many phenomenological uses, to their unified gauge completions in the ultraviolet.
}

\maketitle

\section{Introduction}
\label{sec:intro}

There are many phenomenological reasons to entertain an extension of the Standard Model (SM) gauge algebra $\sm:=\su(3)\oplus \su(2)_L\oplus \u_Y$ by $\u_X$ summands which, after being spontaneously broken, would give rise to neutral $\ZP$ gauge bosons. For example, if weakly coupled to the SM fields, a $\ZP$ boson could mediate interactions with a dark sector. If light ($m_X \lesssim 4$ GeV~\cite{Greljo:2022dwn}) and suitably coupled to leptons, a $\ZP$ can mediate a 1-loop contribution to the anomalous magnetic moment of the muon that resolves an estimated 4.2$\sigma$  discrepancy~\cite{Muong-2:2021ojo,Aoyama:2020ynm} between data and the SM prediction. 
If heavier ($m_X \gtrsim 2$ TeV, say) and equipped with quark flavour violating and lepton flavour universality violating interactions, a $\ZP$ can resolve a collection of measurements~\cite{LHCb:2020lmf,LHCb:2020gog,LHCb:2020zud,LHCb:2021awg,LHCb:2021vsc,LHCb:2014cxe,LHCb:2015wdu,LHCb:2016ykl,LHCb:2021zwz,LHCb:2017avl,LHCb:2021trn} (the `$b\to s\mu\mu$ anomalies') in semi-leptonic $B$-meson decays that are in tension with the SM. Finally, the $\u_X$ gauge symmetry itself can be put to good phenomenological use, for example in explaining the flavour puzzle \`a la Froggatt and Nielsen~\cite{Froggatt:1978nt}. 

To make life (or at least model-building) simpler, it is a good idea to ensure the $\u_X$ extension of the SM gauge algebra is free of perturbative gauge anomalies.\footnote{\label{foot:su2}There are no non-perturbative anomalies beyond Witten's $\SUL$ anomaly in SM$\times \U_X$ extensions~\cite{Davighi:2019rcd}.} 
If this is not the case, the UV completion of the model must feature extra chiral fermions that restore anomaly cancellation. 
If the extra fermions are chiral under $\sm$, then it will be difficult to give them big enough masses to have eluded discovery at the LHC.
On the other hand, if the extra fermions are SM singlets, then their masses can comfortably reside at the heavy scale of $\u_X$ breaking, so there is no tension here with collider bounds.\footnote{There are, of course, other examples of anomalous $\u_X$ models which are phenomenologically viable, that need SM-charged extra fermions to cancel anomalies. Notably, if the extra fermions are {\em vector-like} under SM, as in {\em e.g.}~\cite{Davighi:2021oel,DiLuzio:2022ziu}, then there is no problem with their being heavy.} Thus, the simplest strategies for building a consistent $\sm \oplus \u_X$ model are to ensure gauge anomaly cancellation amongst the SM fermions on their own, or allowing some number $N$ of SM singlets. In the cases $N \leq 3$, {\em a.k.a.} the `SM$+3\nu_R$', the complete space of flavour non-universal anomaly-free $\u_X$ extensions of the SM has been numerically scanned in Ref.~\cite{Allanach:2018vjg}, and even parametrised analytically (for the $N=3$ case only) in Ref.~\cite{Allanach:2020zna}.

Despite their many uses, extending the SM by an anomaly-free $\u_X$ gauge symmetry goes against the popular idea of unification, whereby one asks for fewer forces not more, or at least fewer fields, at high energies. There is not a huge variety of unified gauge theories in which one can embed the SM on its own, given in particular its intricate pattern of hypercharge quantum numbers, and this is clearly made more difficult by seeking to embed a second set of $\u_X$ charges also. To illustrate the point, if there were a single generation of SM fermions, there is no additional $\u_X$ whatsoever that could be embedded in the $\su(5)$ GUT~\cite{Georgi:1974sy}, while there is an unique option that embeds inside $\so(10)$~\cite{Fritzsch:1974nn,georgi1975particles}, namely $X=B-L$. Of course, there is not one generation but three in Nature, and this opens up a much wider arena of flavour {\em non}-universal gauge models, to which we return our attention shortly.

Extending the SM by $\u_X$ also exacerbates the issue of the Landau pole in the SM; now there are {\em two} $\U$ gauge symmetries, each of which becomes more strongly coupled at higher energy scales. 
Even though we got lucky with hypercharge, for which the Landau pole is at a very high scale and so can perhaps be waved away by appealing to quantum gravity effects, this is not necessarily the case for a phenomenologically-motivated $\u_X$. It was recently shown~\cite{Bause:2021prv} that $\ZP$ models for the $b\to s\mu\mu$ anomalies suffer from sub-Planckian Landau poles, that can be as low as 100 TeV for realistic models. This means that $\ZP$ models for the $b\to s\mu\mu$ anomalies require new physics of some kind to tame this running -- in Ref.~\cite{Bause:2021prv} extra scalars and fermions were included to soften the running of the $\u_X$ gauge coupling. Arguably, the resulting theories are starting to look even more complicated at higher energies, taking us further from the ideals of unification.

The other option for curing a low scale Landau pole is to try to reverse the direction of its running through self-interacting gluon contributions; in other words, to ameliorate the Landau pole by embedding the $\u_X$ inside a semi-simple gauge algebra, and taking seriously the idea of unification. But, given any old anomaly-free $\u_X$ extension of the SM, it is usually not obvious whether the $\sm\oplus \u_X$ theory can be embedded in a semi-simple gauge algebra, nor is it so obvious what strategy to employ to find this algebra. 

If one restricts the matter content of the UV model, however, the question at least becomes sharply posed, because the number of possible Lie algebras $\g$ in which $\sm \oplus \u_X$ embeds becomes finite.
Recently, Ref.~\cite{Allanach:2021bfe} classified all possible semi-simple gauge extensions of $\sm$ that do not require extra fermions beyond those of the SM+$3\nu_R$, and that are free of perturbative gauge anomalies.\footnote{Unlike the case of a $\U_X$ extension of the SM gauge group (footnote \ref{foot:su2}), for some of the semi-simple $\g$ in the list of~\cite{Allanach:2021bfe} there {\em are} potentially non-perturbative (or `global') gauge anomalies associated with the corresponding gauge groups $G$. These are associated to $SU(2)$ factors in $G$, or, more generally, to $Sp(2N)$ factors for various $N$. We return to the issue of global anomaly cancellation in Section \ref{sec:global}.}
The result was a finite set, call it $\S$, of 340 inequivalent Lie algebras $\g$. The richness of this list is in large part a result of there being three families of SM fermions. This allows for options in which gauge and flavour symmetries are unified that have been little studied until now; for example, $\g=\su(4)\oplus \sp(6)_L\oplus\sp(6)_R$ in which electroweak and flavour symmetries are unified (a model that was explored in detail in~\cite{Davighi:2022fer}), or $\g=\su(12)\oplus \su(2)_L\oplus \su(2)_R$ in which colour and flavour are unified.

Equipped with the list $\S$, one can find all possible {anomaly-free} $\sm\oplus\u_X$ subgroups, in fact all possible subgroups of the more general form $\sm \oplus \u_{X^1} \oplus \u_{X^2} \oplus \dots$,\footnote{Such multiple $\u$ extensions of the SM have been used \emph{e.g.} in neutrino mass model building~\cite{Froggatt:1996np,Froggatt:1998he}.} contained in any of the algebras $\g \in \S$.
In this paper, we have two related goals:
\begin{enumerate}
\item To parametrise the space of all {anomaly-free} abelian extensions of the SM, by any number of $\u$'s, that embed in {anomaly-free} semi-simple algebras $\g \in \S$.
\item To provide a simple way of testing whether a given {anomaly-free} abelian extension of the SM sits in one of the {anomaly-free} semi-simple algebras $\g \in \S$.
\end{enumerate}
After achieving the first goal, the second goal follows swiftly.
To achieve goal 1., our method is straightforward: for a sufficiently large subset of algebras $\g$ in $\S$, we first compute the centraliser $\mathfrak{C}_{\g}(\sm)$ of $\sm$ in $\g$, then find the Cartan subalgebra $\h_{\mathfrak{C}_{\g}(\sm)}$ thereof, from which we can extract the most general sets of charges that can be embedded in $\g$.

Unlike the space of anomaly-free $\u_X$ extensions of $\sm$, which is some complicated set of rational points cut out by the intersection of a quadratic and a cubic equation, the space of anomaly-free abelian extensions with semi-simple completions is a `nice' linear space, being a union of planes. For want of a better name, at times we refer to this space as `Flatland'~\cite{abbott1885flatland}.\footnote{A more accurate (but less memorable) name might have been `Linear-land'.} 

\sloppy To give an explicit example of one of these component planes, the set of {$\u_{X^1} \oplus \u_{X^2} \oplus \dots$  charge assignments} that embeds in the flavour non-universal algebra $\g_2=\so(10)_1 \oplus \so(10)_2 \oplus \so(10)_3$ is the plane
$P_2 = \mathrm{Span}(Y_1,Y_2,Y_3,BL_1,BL_2,BL_3)$, where $BL$ is an abbreviation for $B-L$.
To give another simple example, for the colour-flavour unification group $\g_6:=\su(12)\oplus \su(2)\oplus \su(2)$, the corresponding plane is
$P_6=\mathrm{Span}(Y,BL,B_{12},B_{23},L_{12},L_{23})$, where $B_{12}$ is an abbreviation for $B_1-B_2$. Equivalently, the plane $P_6$ is, up to hypercharge, the space of traceless combinations of $B_1$, $B_2$, $B_3$, $L_1$, $L_2$, and $L_3$, which coincides precisely with the space of anomaly-free {abelian extensions} with {\em vector-like} charges (see Section~\ref{sec:VL}). The fact that such vector-like $\u_X$ models embed in the colour-flavour unification algebra $\su(12)\oplus \su(2)\oplus \su(2)$ has already been put to phenomenological use in Ref.~\cite{Davighi:2022qgb}, {which reveals a novel connection between flavour non-universality and the stability of the proton}. 

We find that a total of 8 such planes, like $P_2$ and $P_6$ just described, are needed to parametrise Flatland, the space of all possible {$\sm \oplus \u_{X^1} \oplus \u_{X^2} \oplus \dots$ theories} with semi-simple completions, taking care to ensure there is a possible gauge {\em group} that is free of not just local anomalies, but also global anomalies.\footnote{For completeness, we also present results in the case that one ignores global anomalies.}
For some of these planes, we show how to derive predicates ({\em i.e.} true-or-false statements) that test whether {given sets $X^n$ of charges} lie in that plane. Such a predicate must take into account the freedom to permute family indices of each species of SM fermion, and some planes in Flatland are difficult to characterise by a simple permutation-invariant predicate. Nonetheless, by cycling through family permutations on a computer, one can straightforwardly determine whether a given abelian extension by $\u_{X^1} \oplus \u_{X^2} \oplus \dots$ sits in any of these 8 planes, and thence identify its possible semi-simple completions.\footnote{Technically, the program does not tell us every possible semi-simple $\g$ into which a given $\sm \oplus \u_{X^1} \oplus \u_{X^2} \oplus \dots$ theory embeds, but only a set of `maximal' such $\g$ (where an algebra $\g$ is maximal in $\S$ if there is no other $\g^\prime \in \S$ in which $\g$ embeds - see Section~\ref{sec:formalism}).}

We write a computer program that performs these tasks, and share the code with the \texttt{arXiv} submission of this article. We hope that this provides a useful tool when model-building with {(multiple)} $\u$ extensions of the SM, by pointing the way towards the possible unified gauge models in the UV.

We apply our results to many anomaly-free $\u_X$ models from the literature. 
As mentioned, any vector-like anomaly-free $\u_X$, such as $B-L$, $B_3-L_2$, $B-3L_\mu$, or $L_\mu-L_\tau$, necessarily embeds (at least) in the colour-flavour unification algebra $\g_6$. It is therefore for the chiral $\u_X$ models that the situation is more interesting. We give a summary of results in Table~\ref{tab:models2}. 
To give two more examples, we find that the `$DY_3$ model' of~\cite{Allanach:2019iiy,Allanach:2021kzj}, which is the unique anomaly-free $\u_X$ model in which a single family of quarks and two families of leptons are charged, does not sit in any semi-simple $\g$. Nor do any of the 21 chiral `muoquark models' from~\cite{Greljo:2021npi,Greljo:2022dwn} have semi-simple completions.

Finally, we ask and answer the quantitative question of `how many' anomaly-free $\sm\oplus \u_X$ models can be embedded in semi-simple gauge theories. By scanning through the `anomaly-free atlas' of Ref.~\cite{Allanach:2018vjg}, we find that roughly $2.5\%$ of solutions in the anomaly-free atlas with a maximum charge of 10 have semi-simple completions. 
This fraction falls roughly exponentially with the maximum charge (see Fig.~\ref{fig:AtlasSummary}). 

The rest of the paper is structured as follows.
In Section~\ref{sec:formalism} {we introduce the notion of an abelian extension of the SM, setting out the definitions and formalism we shall need in the paper.}
In Section~\ref{sec:detail} we describe our method {for finding which abelian extensions have semi-simple completions,} in detail: this Section can be skipped without loss of continuity, if one is only interested in our results. In Section~\ref{sec:results} we find the collection of planes {parametrising all possible abelian extensions that have semi-simple completions.} We apply these results to a selection of explicit $\u_X$ models from the literature, and to all solutions in the `anomaly-free atlas' of~\cite{Allanach:2018vjg}. Finally, in Section~\ref{sec:program} we explain how to use the short computer program \texttt{Test\_your\_own\_charges.nb}, included with the \texttt{arXiv} submission of this article, with which the user can test whether an input {$\sm \oplus \u_{X^1} \oplus \u_{X^2} \oplus \dots$ theory} has a semi-simple completion (and if so, find out in which of the 8 planes of Flatland it sits). We also itemise a selection of other computer programs that are included with the article.

\section{Abelian extensions from semi-simple algebras} \label{sec:formalism}

Let $\sm:=\su(3)\oplus \su(2)_L\oplus \u_Y$ denote the gauge algebra of the SM, and let $\gamma:\sm \to \su(48)$ denote an embedding that defines the representation of the 48 Weyl fermions of the SM+$3\nu_R$. {To set our conventions, the SM Weyl fermions are in the following representations of $\sm$, where $j\in\{1,2,3\}$ is the family index: $q_j \sim {\bf (3,2)}_1$, $u_j^c \sim {\bf (\overline{3},1)}_{-4}$,  $d_j^c \sim {\bf (\overline{3},1)}_{2}$, 
$\ell_j \sim {\bf (1,2)}_{-3}$, $e_j^c \sim {\bf (1,1)}_{6}$,  and $\nu_j^c \sim {\bf (1,1)}_{0}$. All these fields are left-handed Weyl fermions.
}

Ref.~\cite{Allanach:2021bfe} recently catalogued all possible semi-simple gauge extensions of the SM+$3\nu_R$; roughly, this means finding all
(complex) semi-simple Lie algebras $\g$ such that there is a pair of embeddings $\alpha:\sm\to \g$ and $\beta:\g \to \su(48)$ that are compatible with $\gamma$, in the sense that the diagram 
\begin{equation} \label{eq:triangle}
\begin{tikzcd}
& \g \arrow{rd}{\beta} &\\
\sm \arrow{rr}{\gamma}\arrow{ur}{\alpha} && \su(48)
\end{tikzcd}
\end{equation}
commutes, and such that $\g$ is moreover free of local gauge anomalies.
Two triples are said to be equivalent $(\g,\alpha,\beta)\sim(\g^\prime,\alpha^\prime,\beta^\prime)$ if there is an inner automorphism $j$ of $\su(48)$ and an automorphism $O$ of $\g$ such that $(\g,O\circ \alpha, j\circ \beta \circ O^{-1})=(\g^\prime,\alpha^\prime,\beta^\prime)$. A grand total of 340 such inequivalent $(\g,\alpha,\beta)$ were found. Let us call this set of extensions $\S$ (for `semi-simple'). 

Of these algebras, 24 are maximal, meaning there is no inequivalent $(\g^\prime,\alpha^\prime,\beta^\prime)$ such that there are embeddings $j:\g\rightarrow \g^\prime$ and $i:\su(48)\rightarrow \su(48)$ with $j\circ \alpha=\alpha^\prime$ and $\beta^\prime \circ j=i\circ \beta$. {These 24 maximal algebras, as taken from~\cite{Allanach:2021bfe}, are reproduced in the first 24 rows of Table~\ref{tab:maximal}.}
We return to the issue of maximal algebras in the next Subsection.

Let $\sm_S:=\su(3)\oplus \su(2)_L$ be the semi-simple part of the SM gauge algebra. 
An abelian extension of the SM is a representation 
\begin{align}
\epsilon: \sm_S\oplus \mathfrak{a}\rightarrow \su(48),
\end{align} where $\mathfrak{a}$ is abelian, and where there is an embedding 
\be \label{eq:iota}
\iota:\sm\rightarrow \sm_S\oplus \mathfrak{a}
\ee 
such that $\epsilon\circ \iota$ is the usual SM fermionic representation $\gamma$. It is preferable to include $\u_Y$ in the abelian part $\a$ because of the freedom to do linear field redefinitions on the abelian gauge fields (at the expense of introducing kinetic mixing). Including $\u_Y$ in the abelian part thus allows us to properly account for all possible extensions.
We restrict our attention to those extensions where \begin{align}\label{eq:aActsReph}
\epsilon(\h_{\sm_S}\oplus\a)\subset \h_{\su(48)},
\end{align} for some choice of Cartan subalgebras, since these are the ones which act on fermions via a phase.\footnote{{
This condition would exclude embeddings analogous to $\u\rightarrow \su(2):X\mapsto J_{\pm}$, where $J_\pm= \sigma_1 \pm i \sigma_2$ are the usual ladder operators. The unpleasantry of such an embedding is that $J_\pm$ is not diagonalisable (\emph{i.e.} is not a semi-simple element), and it will not act reducibly on the SM fermions.}
} The same information contained in the embedding $\epsilon:\sm\rightarrow \su(48)$ can be expressed, less formally (but perhaps more familiarly), by a particular
\be
\sm\oplus \u_{X^1}\oplus \u_{X^2}\oplus \cdots \u_{X^N}\,,
\ee
for $N$ specific (independent) sets of charge assignments $X^n \in \Q^{18}$ (where a  point in $\Q^{18}$ records the charges of the 18 fermions of SM+$3\nu_R$).

In this work, we will find all possible abelian extensions of the SM+$3\nu_R$ 
that can be embedded into at least one of the anomaly-free semi-simple gauge models in $\S$. Namely, we find abelian extensions such that there exists a $(\g,\alpha,\beta)$ and an  embedding
\begin{align}
\tau: \sm_S \oplus \mathfrak{a}\rightarrow \g
\end{align} such that  $\tau|_\sm=\alpha$ and the diagram 
\begin{equation} \label{eq:triangleAbelien}
\begin{tikzcd}
& \g \arrow{rd}{\beta} &\\
\sm_S\oplus \a \arrow{rr}{i\circ \epsilon}\arrow{ur}{\tau} && \su(48)
\end{tikzcd}
\end{equation}
commutes, where $i:\su(48)\rightarrow \su(48)$ is an inner automorphism (which takes account of \emph{e.g.} family permutations).

Clearly, if an abelian extension embeds into any of the 340 algebras in $\S$ then it must also embed into a maximal one. Hence we can restrict our attention to maximal $(\g,\alpha,\beta)$.

Since $\a$ commutes with $\sm=\sm_S \oplus \u_Y$, $\tau(\a)$ must be a subalgebra of the centraliser $\mathfrak{C}_{\g}(\sm)$ of $\sm$ in $\g$, which is defined to be
\begin{align}
\label{eq:Cg_def}
\mathfrak{C}_\g(\sm) = \{ v \in \g \, |\, [v, u] = 0 \, \forall u \in \alpha(\sm) \}.
\end{align}
{More formally, we denote the algebra itself $\mathfrak{C}_\g(\sm)$ and its embedding into $\g$ as $\rho:\mathfrak{C}_\g(\sm)\rightarrow \g$.}  Using standard theorems in Lie algebra theory (see {\em e.g.}~\cite{jacobson1989completely}), it is easy to show that $\mathfrak{C}_{\g}(\sm)$ is a reductive Lie algebra, where recall that a reductive Lie algebra is one that can be written as a direct sum of simple factors and $\u$ factors.

Given a chosen Cartan subalgebra $\h_{\mathfrak{C}_{\g}(\sm)}$ of $\mathfrak{C}_{\g}(\sm)$ we have a diagram
\begin{equation}\label{eq:diaCart}
\begin{tikzcd}
& \g \arrow{rd}{\beta} &\\
\sm_S\oplus \h_{\mathfrak{C}_{\g}(\sm)} \arrow{rr}{\epsilon_{\mathfrak{C}}}\arrow{ur}{\tau_{\mathfrak{C}}} && \su(48)
\end{tikzcd}
\end{equation}
where $\tau_{\mathfrak{C}}$ is induced by the embedding of $\h_{\mathfrak{C}_{\g}(\sm)}$ into $\g$ and  $\epsilon_{\mathfrak{C}}$ can be defined as the unique embedding such that this diagram commutes. The important property of Diagram~(\ref{eq:diaCart}) is that it is maximal with respect to the Diagrams in~(\ref{eq:triangleAbelien}). By `maximal' we mean that, for any valid $\epsilon:\sm_S\oplus \a\rightarrow \su(48)$ satisfying (\ref{eq:triangleAbelien}), there is an embedding $t:\mathfrak{sm}_S\oplus\mathfrak{a}\rightarrow \mathfrak{sm}_S\oplus\mathfrak{h}_{\mathfrak{C}_{\mathfrak{g}}}$ such that there exists a commutative diagram
\begin{equation}
\begin{tikzcd}
	& {\mathfrak{g}} \\
	{\mathfrak{sm}_S\oplus\mathfrak{h}_{\mathfrak{C}_{\mathfrak{g}}(\mathfrak{sm})}} && {\mathfrak{su}(48)} \\
	& {\mathfrak{g}} \\
	{\mathfrak{sm}_S\oplus\mathfrak{a}} && {\mathfrak{su}(48)}
	\arrow["{\tau_{\mathfrak{C}}}", from=2-1, to=1-2]
	\arrow["\beta", from=1-2, to=2-3]
	\arrow["{\epsilon_{\mathfrak{C}}}",from=2-1, to=2-3]
	\arrow["{i \circ \epsilon}"', from=4-1, to=4-3]
	\arrow["\beta", from=3-2, to=4-3]
	\arrow["\tau", from=4-1, to=3-2]
	\arrow["t"{description}, color={rgb,255:red,214;green,92;blue,92}, curve={height=-6pt}, from=4-1, to=2-1]
	\arrow["{i_1}"{description, pos=0.3}, color={rgb,255:red,214;green,92;blue,92}, curve={height=-6pt}, from=3-2, to=1-2]
	\arrow["{i_2}"{description}, color={rgb,255:red,214;green,92;blue,92}, curve={height=-6pt}, from=4-3, to=2-3]
\end{tikzcd}
\end{equation} 
where all morphisms are embeddings and $i_1$ and $i_2$ are inner automorphisms. In particular, commutativity of the `left face' implies the maps $\tau$ and $\tau_{\mathfrak{C}}\circ t$ agree up to an inner automorphism $i_1$ on $\g$; colloquially, we would say that $\tau$ `factors through' $\sm_S \oplus \h_{\C_{\g}}$. Note that the choice of the specific $\h_{\mathfrak{C}_{\g}(\sm)}$ does not matter, since all choices are related by inner automorphism.

{
The map $\beta:\g\to \su(48)$ can then be used to determine how $\h_{\mathfrak{C}_{\g}(\sm)}$  acts on the fermions, via the weight system. This can be translated into a set of $\u_X$ charges, which span a plane $P_{\g}$ in the rational `charge space' $\mathbb{Q}^{18}$. The question of whether some abelian extension $\epsilon: \sm_S\oplus \a\rightarrow \su(48)$ sits in any $(\g,\alpha,\beta)$  then reduces to the question of asking whether the charges associated with $\a$ sit inside the (maximal) planes $P_{\g}$ (up to family permutations that correspond to relabeling).

}

Our procedure can thus be summarised as follows: for each of the maximal algebras $(\g,\alpha,\beta)$ in $\S$, we 
\begin{enumerate}
\item find the centraliser $\mathfrak{C}_{\g}(\sm)$, 
\item find a Cartan subalgebra $\h_{\mathfrak{C}_{\g}(\sm)}$, 
\item extract the weights associated to the map $\beta$, 
\item thence find a basis of independent charge vectors, thence the plane $P_{\g}$ that they span.
\end{enumerate}
This procedure, which is detailed in the next Section, can already be made fairly `algorithmic'. It can, in fact, be carried out for all 340 algebras in $\S$.

\begin{table}
\centerline{
\begin{scriptsize}
\begin{tabular}{|c|c|c|}
\hline
$i$ & \bf{Maximal gauge algebra} $\g_i$ & \bf{Fermion representations corresponding to $\beta$} \\ \hline 1 &  $\mathfrak{so}(10) \oplus \mathfrak{su}(2)$ & $({\mathbf{16}}, {\mathbf{3}})$
  \\ \hline 2 &  $\mathfrak{so}(10)^{\oplus 3} $ & $({\mathbf{16}}, {\mathbf{1}}, {\mathbf{1}}) \oplus ({\mathbf{1}}, {\mathbf{16}}, {\mathbf{1}}) \oplus ({\mathbf{1}}, {\mathbf{1}}, {\mathbf{16}})$
  \\ \hline 3 &  $\mathfrak{so}(10)^{\oplus 2}  \oplus \mathfrak{su}(2)$ & $({\mathbf{16}}, {\mathbf{1}}, {\mathbf{1}}) \oplus ({\mathbf{1}}, {\mathbf{16}}, {\mathbf{2}})$
  \\ \hline 4 &  $\mathfrak{su}(4) \oplus \mathfrak{sp}(6)^{\oplus 2} $ & $(\overline{\mathbf{4}}, {\mathbf{6}}, {\mathbf{1}}) \oplus ({\mathbf{4}}, {\mathbf{1}}, {\mathbf{6}})$
  \\ \hline 5 &  $\mathfrak{su}(4)^{\oplus 2}  \oplus \mathfrak{sp}(6)$ & $(\overline{\mathbf{4}}, {\mathbf{6}}, {\mathbf{1}}) \oplus ({\mathbf{4}}, {\mathbf{1}}, {\mathbf{6}})$
  \\ \hline 6 &  $\mathfrak{su}(12) \oplus \mathfrak{su}(2)^{\oplus 2} $ & $(\overline{\mathbf{12}}, {\mathbf{2}}, {\mathbf{1}}) \oplus ({\mathbf{12}}, {\mathbf{1}}, {\mathbf{2}})$
  \\ \hline 7 &  $\mathfrak{su}(4) \oplus \mathfrak{sp}(4)^{\oplus 2}  \oplus \mathfrak{so}(10)$ & $(\overline{\mathbf{4}}, {\mathbf{4}}, {\mathbf{1}}, {\mathbf{1}}) \oplus ({\mathbf{4}}, {\mathbf{1}}, {\mathbf{4}}, {\mathbf{1}}) \oplus ({\mathbf{1}}, {\mathbf{1}}, {\mathbf{1}}, {\mathbf{16}})$
  \\ \hline \textcolor{myred}{8$^\ast$} &  \textcolor{myred}{$\mathfrak{su}(5) \oplus \mathfrak{su}(2)^{\oplus 3} $} & \textcolor{myred}{$(\overline{\mathbf{5}}, {\mathbf{3}}, {\mathbf{1}}, {\mathbf{1}}) \oplus ({\mathbf{10}}, {\mathbf{1}}, {\mathbf{3}}, {\mathbf{1}}) \oplus ({\mathbf{1}}, {\mathbf{1}}, {\mathbf{1}}, {\mathbf{2}}) \oplus (\mathbf{1},\mathbf{1},\mathbf{1},\mathbf{1})$}
  \\ \hline 9 &  $\mathfrak{su}(5) \oplus \mathfrak{su}(2)^{\oplus 3} $ & $(\overline{\mathbf{5}}, {\mathbf{3}}, {\mathbf{1}}, {\mathbf{1}}) \oplus ({\mathbf{10}}, {\mathbf{1}}, {\mathbf{3}}, {\mathbf{1}}) \oplus ({\mathbf{1}}, {\mathbf{1}}, {\mathbf{1}}, {\mathbf{3}})$
  \\ \hline \textcolor{myred}{10$^\ast$} &  \textcolor{myred}{$\mathfrak{su}(5) \oplus \mathfrak{su}(2)^{\oplus 3} $} & \textcolor{myred}{$(\overline{\mathbf{5}}, {\mathbf{1}}, {\mathbf{1}}, {\mathbf{1}}) \oplus (\overline{\mathbf{5}}, {\mathbf{2}}, {\mathbf{1}}, {\mathbf{1}}) \oplus ({\mathbf{10}}, {\mathbf{1}}, {\mathbf{3}}, {\mathbf{1}}) \oplus ({\mathbf{1}}, {\mathbf{1}}, {\mathbf{1}}, {\mathbf{2}}) \oplus (\mathbf{1},\mathbf{1},\mathbf{1},\mathbf{1})$}
  \\ \hline \textcolor{myred}{11$^\ast$} &  \textcolor{myred}{$\mathfrak{su}(5) \oplus \mathfrak{su}(2)^{\oplus 3} $} & \textcolor{myred}{$(\overline{\mathbf{5}}, {\mathbf{1}}, {\mathbf{1}}, {\mathbf{1}}) \oplus (\overline{\mathbf{5}}, {\mathbf{2}}, {\mathbf{1}}, {\mathbf{1}}) \oplus ({\mathbf{10}}, {\mathbf{1}}, {\mathbf{3}}, {\mathbf{1}}) \oplus ({\mathbf{1}}, {\mathbf{1}}, {\mathbf{1}}, {\mathbf{3}})$}
  \\ \hline \textcolor{myred}{12$^\ast$} &  \textcolor{myred}{$\mathfrak{su}(5) \oplus \mathfrak{su}(2)^{\oplus 3} $} & \textcolor{myred}{$({\mathbf{10}}, {\mathbf{1}}, {\mathbf{1}}, {\mathbf{1}}) \oplus (\overline{\mathbf{5}}, {\mathbf{3}}, {\mathbf{1}}, {\mathbf{1}}) \oplus ({\mathbf{10}}, {\mathbf{1}}, {\mathbf{2}}, {\mathbf{1}}) \oplus ({\mathbf{1}}, {\mathbf{1}}, {\mathbf{1}}, {\mathbf{2}}) \oplus (\mathbf{1},\mathbf{1},\mathbf{1},\mathbf{1})$}
  \\ \hline 13 &  $\mathfrak{su}(5) \oplus \mathfrak{su}(2)^{\oplus 3} $ & $({\mathbf{10}}, {\mathbf{1}}, {\mathbf{1}}, {\mathbf{1}}) \oplus (\overline{\mathbf{5}}, {\mathbf{3}}, {\mathbf{1}}, {\mathbf{1}}) \oplus ({\mathbf{10}}, {\mathbf{1}}, {\mathbf{2}}, {\mathbf{1}}) \oplus ({\mathbf{1}}, {\mathbf{1}}, {\mathbf{1}}, {\mathbf{3}})$
  \\ \hline \textcolor{myred}{14$^\ast$} &  \textcolor{myred}{$\mathfrak{su}(5)^{\oplus 2}  \oplus \mathfrak{so}(10) \oplus \mathfrak{su}(2)$} & \textcolor{myred}{$(\overline{\mathbf{5}}, {\mathbf{1}}, {\mathbf{1}}, {\mathbf{1}}) \oplus ({\mathbf{10}}, {\mathbf{1}}, {\mathbf{1}}, {\mathbf{1}}) \oplus ({\mathbf{1}}, \overline{\mathbf{5}}, {\mathbf{1}}, {\mathbf{1}}) \oplus ({\mathbf{1}}, {\mathbf{10}}, {\mathbf{1}}, {\mathbf{1}}) \oplus ({\mathbf{1}}, {\mathbf{1}}, {\mathbf{16}}, {\mathbf{1}}) \oplus ({\mathbf{1}}, {\mathbf{1}}, {\mathbf{1}}, {\mathbf{2}})$}
  \\ \hline 15 &  $\mathfrak{su}(5)^{\oplus 3}  \oplus \mathfrak{su}(2)$ & $(\overline{\mathbf{5}}, {\mathbf{1}}, {\mathbf{1}}, {\mathbf{1}}) \oplus ({\mathbf{10}}, {\mathbf{1}}, {\mathbf{1}}, {\mathbf{1}}) \oplus ({\mathbf{1}}, \overline{\mathbf{5}}, {\mathbf{1}}, {\mathbf{1}}) \oplus ({\mathbf{1}}, {\mathbf{10}}, {\mathbf{1}}, {\mathbf{1}}) \oplus ({\mathbf{1}}, {\mathbf{1}}, \overline{\mathbf{5}}, {\mathbf{1}}) \oplus ({\mathbf{1}}, {\mathbf{1}}, {\mathbf{10}}, {\mathbf{1}}) \oplus ({\mathbf{1}}, {\mathbf{1}}, {\mathbf{1}}, {\mathbf{3}})$
  \\ \hline 16 &  $\mathfrak{su}(8) \oplus \mathfrak{so}(10) \oplus \mathfrak{su}(2)^{\oplus 2} $ & $({\mathbf{1}}, {\mathbf{16}}, {\mathbf{1}}, {\mathbf{1}}) \oplus (\overline{\mathbf{8}}, {\mathbf{1}}, {\mathbf{2}}, {\mathbf{1}}) \oplus ({\mathbf{8}}, {\mathbf{1}}, {\mathbf{1}}, {\mathbf{2}})$
  \\ \hline 17 &  $\mathfrak{su}(4) \oplus \mathfrak{sp}(4) \oplus \mathfrak{so}(10) \oplus \mathfrak{su}(2)^{\oplus 2} $ & $(\overline{\mathbf{4}}, {\mathbf{4}}, {\mathbf{1}}, {\mathbf{1}}, {\mathbf{1}}) \oplus ({\mathbf{1}}, {\mathbf{1}}, {\mathbf{16}}, {\mathbf{1}}, {\mathbf{1}}) \oplus ({\mathbf{4}}, {\mathbf{1}}, {\mathbf{1}}, {\mathbf{2}}, {\mathbf{2}})$
  \\ \hline 18 &  $\mathfrak{su}(4) \oplus \mathfrak{sp}(4) \oplus \mathfrak{so}(10) \oplus \mathfrak{su}(2)^{\oplus 2} $ & $(\overline{\mathbf{4}}, {\mathbf{4}}, {\mathbf{1}}, {\mathbf{1}}, {\mathbf{1}}) \oplus ({\mathbf{1}}, {\mathbf{1}}, {\mathbf{16}}, {\mathbf{1}}, {\mathbf{1}}) \oplus ({\mathbf{4}}, {\mathbf{1}}, {\mathbf{1}}, {\mathbf{2}}, {\mathbf{2}})$
  \\ \hline 19 &  $\mathfrak{su}(4) \oplus \mathfrak{sp}(6) \oplus \mathfrak{su}(2)^{\oplus 3} $ & $(\overline{\mathbf{4}}, {\mathbf{6}}, {\mathbf{1}}, {\mathbf{1}}, {\mathbf{1}}) \oplus ({\mathbf{4}}, {\mathbf{1}}, {\mathbf{2}}, {\mathbf{2}}, {\mathbf{1}}) \oplus ({\mathbf{4}}, {\mathbf{1}}, {\mathbf{1}}, {\mathbf{1}}, {\mathbf{2}})$
  \\ \hline 20 &  $\mathfrak{su}(4) \oplus \mathfrak{sp}(6) \oplus \mathfrak{su}(2)^{\oplus 3} $ & $(\overline{\mathbf{4}}, {\mathbf{6}}, {\mathbf{1}}, {\mathbf{1}}, {\mathbf{1}}) \oplus ({\mathbf{4}}, {\mathbf{1}}, {\mathbf{2}}, {\mathbf{2}}, {\mathbf{1}}) \oplus ({\mathbf{4}}, {\mathbf{1}}, {\mathbf{1}}, {\mathbf{1}}, {\mathbf{2}})$
  \\ \hline 21 &  $\mathfrak{su}(4)^{\oplus 2}  \oplus \mathfrak{su}(2)^{\oplus 3} $ & $(\overline{\mathbf{4}}, {\mathbf{6}}, {\mathbf{1}}, {\mathbf{1}}, {\mathbf{1}}) \oplus ({\mathbf{4}}, {\mathbf{1}}, {\mathbf{2}}, {\mathbf{2}}, {\mathbf{1}}) \oplus ({\mathbf{4}}, {\mathbf{1}}, {\mathbf{1}}, {\mathbf{1}}, {\mathbf{2}})$
  \\ \hline \textcolor{myred}{22$^\ast$} &  \textcolor{myred}{$\mathfrak{su}(5) \oplus \mathfrak{so}(10) \oplus \mathfrak{su}(2)^{\oplus 3} $} & \textcolor{myred}{$({\mathbf{1}}, {\mathbf{16}}, {\mathbf{1}}, {\mathbf{1}}, {\mathbf{1}}) \oplus (\overline{\mathbf{5}}, {\mathbf{1}}, {\mathbf{2}}, {\mathbf{1}}, {\mathbf{1}}) \oplus ({\mathbf{10}}, {\mathbf{1}}, {\mathbf{1}}, {\mathbf{2}}, {\mathbf{1}}) \oplus ({\mathbf{1}}, {\mathbf{1}}, {\mathbf{1}}, {\mathbf{1}}, {\mathbf{2}})$}
  \\ \hline \textcolor{myred}{23$^\ast$} &  \textcolor{myred}{$\mathfrak{su}(5)^{\oplus 2}  \oplus \mathfrak{su}(2)^{\oplus 3} $} & \textcolor{myred}{$({\mathbf{1}}, \overline{\mathbf{5}}, {\mathbf{1}}, {\mathbf{1}}, {\mathbf{1}}) \oplus ({\mathbf{1}}, {\mathbf{10}}, {\mathbf{1}}, {\mathbf{1}}, {\mathbf{1}}) \oplus (\overline{\mathbf{5}}, {\mathbf{1}}, {\mathbf{2}}, {\mathbf{1}}, {\mathbf{1}}) \oplus ({\mathbf{10}}, {\mathbf{1}}, {\mathbf{1}}, {\mathbf{2}}, {\mathbf{1}}) \oplus ({\mathbf{1}}, {\mathbf{1}}, {\mathbf{1}}, {\mathbf{1}}, {\mathbf{3}})$}
  \\ \hline 24 &  $\mathfrak{su}(4) \oplus \mathfrak{so}(10) \oplus \mathfrak{su}(2)^{\oplus 4} $ & $({\mathbf{1}}, {\mathbf{16}}, {\mathbf{1}}, {\mathbf{1}}, {\mathbf{1}}, {\mathbf{1}}) \oplus ({\mathbf{4}}, {\mathbf{1}}, {\mathbf{2}}, {\mathbf{2}}, {\mathbf{1}}, {\mathbf{1}}) \oplus (\overline{\mathbf{4}}, {\mathbf{1}}, {\mathbf{1}}, {\mathbf{1}}, {\mathbf{2}}, {\mathbf{2}})$
  \\ \hline \hline
  \textcolor{mygreen}{25} & \textcolor{mygreen}{$\mathfrak{su}(5) \oplus \mathfrak{su}(2)^{\oplus 2} $} & \textcolor{mygreen}{$(\overline{\mathbf{5}}, {\mathbf{1}}, {\mathbf{1}}) \oplus (\overline{\mathbf{5}}, {\mathbf{2}}, {\mathbf{1}}) \oplus ({\mathbf{10}}, {\mathbf{1}}, {\mathbf{3}}) \oplus ({\mathbf{1}}, {\mathbf{2}}, {\mathbf{1}}) \oplus (\mathbf{1},\mathbf{1},\mathbf{1})$}\\
  \hline  \textcolor{mygreen}{26 }&  \textcolor{mygreen}{$\mathfrak{su}(5) \oplus \mathfrak{so}(10) \oplus \mathfrak{su}(2)^{\oplus 2} $} &  \textcolor{mygreen}{$({\mathbf{1}}, {\mathbf{16}}, {\mathbf{1}}, {\mathbf{1}}) \oplus (\overline{\mathbf{5}}, {\mathbf{1}}, {\mathbf{2}}, {\mathbf{1}}) \oplus ({\mathbf{10}}, {\mathbf{1}}, {\mathbf{1}}, {\mathbf{2}}) \oplus ({\mathbf{1}}, {\mathbf{1}}, {\mathbf{2}}, {\mathbf{1}})$}\\
  \hline  \textcolor{mygreen}{27} &  \textcolor{mygreen}{$\mathfrak{su}(5)^{\oplus 2}  \oplus \mathfrak{su}(2)^{\oplus 2} $} &  \textcolor{mygreen}{$(\overline{\mathbf{5}}, {\mathbf{1}}, {\mathbf{1}}, {\mathbf{1}})^{\oplus 2}  \oplus ({\mathbf{1}}, \overline{\mathbf{5}}, {\mathbf{1}}, {\mathbf{1}}) \oplus ({\mathbf{1}}, {\mathbf{10}}, {\mathbf{1}}, {\mathbf{1}}) \oplus ({\mathbf{10}}, {\mathbf{1}}, {\mathbf{2}}, {\mathbf{1}}) \oplus ({\mathbf{1}}, {\mathbf{1}}, {\mathbf{1}}, {\mathbf{3}})$}\\
  \hline
  \end{tabular}
  \end{scriptsize}}
\caption{ The complete list of maximal gauge algebra extensions of the SM, assuming no fermion content beyond the SM+$3\nu_R$. {Rows 1--24 are} reproduced from Ref.~\cite{Allanach:2021bfe}. Those algebras which necessarily suffer from global gauge anomalies (in all cases these are associated with odd numbers of $\su(2)$ doublets) are highlighted with an asterisk, and coloured in \textcolor{myred}{red}. Local gauge anomaly cancellation was already taken care of in producing the list of Ref.~\cite{Allanach:2021bfe}. The algebras 25-27, coloured in \textcolor{mygreen}{green}, become maximal only when considering gauge algebras for which there are corresponding gauge groups that are free of both local {\em and global} anomalies {(meaning those rows in red are struck off)}. \label{tab:maximal}}
\end{table}

\subsection{Global anomaly cancellation} \label{sec:global}

As already mentioned, care was taken to ensure that local gauge anomalies cancel for every gauge algebra listed in Ref.~\cite{Allanach:2021bfe}. Since local gauge anomalies can be computed purely from the Lie algebra $\g$ of a gauge group $G$, this condition could be imposed without ambiguity. But of course, it is not enough to ensure that a gauge theory is free of local (perturbative) anomalies, because there may be more subtle global anomalies that are not associated with infinitesimal gauge transformations, which can arise non-perturbatively. The canonical example is the anomaly associated with a single Weyl fermion in the doublet representation of $SU(2)$, in 4d~\cite{Witten:1982fp}. By definition, such global anomalies cannot be computed knowing only the Lie algebra $\g$, but depend on the global structure of the group $G$ itself.

Some of the gauge algebras listed in Ref.~\cite{Allanach:2021bfe} do not correspond to {\em any} anomaly-free gauge theory, irrespective of the freedom to consider gauge groups that differ globally. For example, consider the maximal algebra listed as number 14 in Table~\ref{tab:maximal}, $\g=\su(5)^{\oplus 2} \oplus \so(10) \oplus \su(2)$. All fermion fields are singlets under the $\su(2)$ factor except for a doublet of right-handed neutrinos transforming in the ${\bf (1,1,1,2)}$ representation. The associated gauge group $G$ necessarily features an $SU(2)$ factor, which has a global anomaly~\cite{Witten:1982fp}. 

In a similar way, all the algebras in Table~\ref{tab:maximal} that are highlighted in red and with an asterisk necessarily suffer from global anomalies. Thus, we should remove these from the list of possible $\g$ if we are to consider UV gauge groups that are properly anomaly-free, without any extra fermion content. Of course, after doing this, the resulting cut-down list of algebras is no longer guaranteed to be the complete list of maximal anomaly-free algebras; there may be valid ({\em i.e.} not necessarily anomalous) algebras in $\S$ which embed into one of those maximal algebras we have crossed off, but which do not embed into any algebra remaining in the maximal list. To account for this, one must include three further algebras, listed 25--27 in Table~\ref{tab:maximal}, that become maximal when considering all possible UV gauge models which are free of both local and global anomalies.

In fact, we will present results in two cases: firstly, ignoring the issue of global anomalies (for which we consider the $\g$ labelled 1--24 in Table~\ref{tab:maximal}), and secondly, taking into account the issue of global anomalies in the manner just described (which amounts to crossing off all the `starred' algebras, coloured red in Table~\ref{tab:maximal}, and instead including the algebras 25--27 coloured green.) The result shall be two parametrisations of abelian extensions of the SM that fit in either list of semi-simple completions. In both cases, the parametrisation takes the form of a union of planes, which are different. Clearly, the second set is a subset of the first.

\section{Method} \label{sec:detail}
We now proceed to give the details of the procedure outlined in Section~\ref{sec:formalism}, {for each of the maximal algebras in Table~\ref{tab:maximal}}. 

Before we start, we note that the information contained in Ref.~\cite{Allanach:2021bfe} for each element of $\S$ is not exactly a triple $(\g, \alpha,\beta)$. Instead, it outputs a list of triples consisting of a semi-simple $\g$, a so-called `projection matrix' $\Lambda \alpha$, and a $48$-dimensional representation of $\g$, up to some notion of equivalence. The fact that this set is equivalent to  $\mathbb{S}$ is proven in Appendix B of Ref.~\cite{Tooby-Smith:2021hjk}. 

Since projection matrices feature heavily in this Section, we think it useful to give a self-contained definition, as follows.
Let us start by reviewing the weight space of a reductive Lie algebra $\mathfrak{r}:=\mathfrak{s}\oplus \u^N$,
where $\mathfrak{s}$ is semi-simple. A Cartan subalgebra of $\mathfrak{r}$ is $\mathfrak{h}_\mathfrak{r}=\mathfrak{h}_\mathfrak{s}\oplus \u^N$. The weights then live in the dual of this space, $\mathfrak{h}_\mathfrak{r}^\ast=\mathfrak{h}_\mathfrak{s}^\ast\oplus (\u^N)^\ast$. Given any semi-simple Lie algebra $\mathfrak{s}$, a suitable basis of $\mathfrak{h}_\mathfrak{s}^\ast$ is the `fundamental weights'. Throughout this paper we always write weights in this basis. A basis of $(\u^N)^\ast$ can be formed by taking the dual of a basis $X^n$ of $\u^N$.

Given an embedding $f:\mathfrak{r}\rightarrow \g$ of the reductive Lie algebra $\mathfrak{r}$ into a semi-simple Lie algebra $\g$, we can form the pull-back $\Lambda f: \mathfrak{h}_\g^\ast \rightarrow \mathfrak{h}_\mathfrak{r}^\ast$ which takes $\lambda\in \mathfrak{h}_\g^\ast$ to $\lambda \circ f\in \mathfrak{h}_\mathfrak{r}^\ast$. (Note that in the literature $\Lambda f$ is often restricted to certain subsets of $\mathfrak{h}_\g^\ast$). We will call the operator $\Lambda f$ the projection matrix of $f$. When written in the bases just described, we denote it $[\Lambda f]$.

For the present work, we need not just the projection matrices $\Lambda \alpha$, but the embeddings $\alpha:\sm\to\g$ themselves for each of the maximal algebras $\g_i$ in Table~\ref{tab:maximal}.
The first step of our method is thus to take each projection $\Lambda \alpha_i$ associated to each $\g_i$, and from it deduce a fully fledged embedding $\alpha_i:\sm \to \g_i$.

\subsection{Step 1: Finding the embeddings of $\sm$ in $\g$ } \label{sec:embeddings}

To start, it is helpful to notice that the problem `factorises' in a convenient way. Because each $\g_i$ is semi-simple, one can decompose it uniquely into a direct sum of simple ideals. These decompositions take the form:
\begin{align} \label{eq:gi_decomp}
\g_i= \bigoplus_{\underline{a}\in I_i} \mathfrak{b}_{\underline{a}} \oplus \mathfrak{hor}_i\, ,
\end{align}
where each $\mathfrak{b}_{\ua}$ is a simple ideal for which $\mathrm{Im}(\alpha_i) \cap \mathfrak{b}_{\ua}\neq \emptyset$, and where $\mathfrak{hor}_i$ is the largest (semi-simple) ideal for which $\mathrm{Im}(\alpha_i) \cap \mathfrak{hor}_i=\emptyset$, and $I_i$ is a $\g_i$-specific list that we define shortly. For each $\mathfrak{b}_{\ua}$, there is a non-zero embedding $\alpha_{\ua}:\sm \to \mathfrak{b}_{\ua}$ which can be extracted (as we will do in this Subsection) from a projection matrix $\Lambda \alpha_{\ua}$. From Ref.~\cite{Allanach:2021bfe}, one can see that the 27 maximal gauge algebras in Table~\ref{tab:maximal} are built out of a small set of 12 recurring building blocks $(\mathfrak{b}_{\ua}, \Lambda\alpha_{\ua})$, $\underline{a} \in \{\underline{1},\dots,\underline{12}\}$.\footnote{
It is helpful for us to distinguish two types of label here; an underlined Latin index $\ua$ everywhere refers to a label for a `building block', {\em i.e.} for one of the simple Lie algebras $\mathfrak{b}_{\ua}$ listed in Table~\ref{tab:blocks}, while a Latin index $i$ (without underline) labels the 27 maximal gauge algebras $\g_i$ listed in Table~\ref{tab:maximal}.
} The `building blocks' $(\mathfrak{b}_{\ua}, \Lambda\alpha_{\ua})$, along with other important data that we describe below,
are listed in Table~\ref{tab:blocks}. The list $I_i$ appearing in Eq.~(\ref{eq:gi_decomp}), which may feature repeats, specifies which of these building blocks appear in $\g_i$.

For example, 
\begin{itemize}
\item for $\g_2$ we have $I_2=\{\underline{2},\underline{2},\underline{2}\}$ and $\mathfrak{hor}_{2}=0$, 
\item for $\g_6$ we have $I_6=\{\underline{12}, \underline{7}, \underline{3}\}$ and $\mathfrak{hor}_{6}=0$,
\item for $\g_{27}$ we have $I_{27}=\{\underline{1},\underline{1}\}$ and $\mathfrak{hor}_{27}= \su(2)^{\oplus 2}$.
\end{itemize}

\begin{table}
\begin{center}
\begin{small}
\begin{adjustwidth}{-.65in}{-.65in} 
\begin{tabular}{|c|c|c|c|c|}
\hline
$\ua$ & $\mathfrak{b}_{\underline{a}}$  & $[\,\Lambda \alpha_{\underline{a}}\,]$ & $\mathfrak{C}_{\underline{a}}$ & $[\,\Lambda\rho_{\underline{a}}\,]$
\\ \hline
\underline{1} & $\mathfrak{su}(5)$   & $\left(\begin{smallmatrix} 0 & 0 & 1 & 0\\ 0 & 0 & 0 & 1 \\ 1 & 0 & 0 &0 \\ 3 & 6 & 4 & 2\end{smallmatrix}\right)$ & $\mathfrak{u}(1)_Y$ & $\left(\begin{smallmatrix} 3 & 6 & 4 & 2\end{smallmatrix}\right)$\\ \hline
\underline{2} & $\mathfrak{so}(10)$  & $\left(\begin{smallmatrix} 0 & 0 & 1 & 0 & 0 \\0&  0 & 0 & 0 &1 \\ 1 & 0 & 0 & 0 & 0\\ 3 & 6 & 4 & 0 & 2\end{smallmatrix}\right)$ & $\mathfrak{u}(1)_Y\oplus \mathfrak{u}(1)_{B-L}$  & $\left(\begin{smallmatrix} 3 & 6 & 4 & 0 & 2\\ 0& 0 & -2 & -3 & -1\end{smallmatrix}\right)$ \\ \hline\hline
\underline{3}  & $\mathfrak{su}(2)_R$  & $\left(\begin{smallmatrix} 0 \\ 0 \\ 0 \\ 3\end{smallmatrix} \right)$  & $\mathfrak{u}(1)_{Y-B+L}$ &  $\left(\begin{smallmatrix} 3 \end{smallmatrix} \right)$\\ \hline
\underline{4}  & $\mathfrak{sp}(4)_R$ & $\left(\begin{smallmatrix} 0&0\\ 0&0\\ 0&0\\ 3 & 0\end{smallmatrix}\right)$ & $\mathfrak{su}(2)_{\text{rh}}\oplus \mathfrak{u}(1)_{Y-B+L}$ & $\left(\begin{smallmatrix} 1 & 2 \\3 & 0 \end{smallmatrix} \right)$  \\ \hline
\underline{5}  & $\mathfrak{sp}(6)_R$  &  $\left(\begin{smallmatrix} 0&0&0\\ 0&0&0\\ 0&0&0\\3& 0 & -3\end{smallmatrix}\right)$ & $\mathfrak{su}(3)_{\text{rh}}\oplus \mathfrak{u}(1)_{Y-B+L}$ & $\left(\begin{smallmatrix} 0 & -1 & 0 \\ 1 & 2 & 2 \\ 3 & 0 & -3\end{smallmatrix} \right)$ \\ \hline 
\underline{6}  & $\mathfrak{so}(6)_R$ &  $\left(\begin{smallmatrix} 0& 0& 0 \\ 0& 0& 0  \\ 0& 0& 0  \\ \frac{9}{2} & 3 & \frac{3}{2}\end{smallmatrix} \right)$ & $\mathfrak{su}(3)_{\text{rh}}\oplus \mathfrak{u}(1)_{Y-B+L}$ & $\left(\begin{smallmatrix} 0 & 1 & 0 \\ 0 & 0 & 1\\ \frac{9}{2} & 3 & \frac{3}{2} \end{smallmatrix}\right) $  \\ \hline \hline
\underline{7}  & $\mathfrak{su}(2)_L$  & $\left(\begin{smallmatrix} 0 \\ 0 \\ 1 \\ 0 \end{smallmatrix}\right)$ & $\emptyset$ & $-$ \\ \hline
\underline{8}  & $\mathfrak{sp}(4)_L$  & $\left(\begin{smallmatrix} 0 &0\\ 0 &0\\ 1&2 \\ 0 &0\end{smallmatrix}\right)$ &  $\mathfrak{so}(2)_{\text{lh}}$ &  $\left(\begin{smallmatrix} 1& 0\end{smallmatrix} \right)$ \\ \hline
\underline{9}  & $\mathfrak{sp}(6)_L$ &$\left(\begin{smallmatrix} 0 &0&0\\ 0 &0&0\\ 1&2&3 \\ 0&0&0\end{smallmatrix}\right)$  & $\mathfrak{so}(3)_{\text{lh}}$ &   $\left(\begin{smallmatrix} 0 & 2 & 0 \end{smallmatrix}\right)$\\ \hline \hline
\underline{10}  & $\mathfrak{su}(4)_{ps}$ &  $\left(\begin{smallmatrix} 0&1&0\\ 1&0&0\\ 0&0&0\\-1&-2&-3\end{smallmatrix}\right)$  & $\mathfrak{u}(1)_{B-L}$ &  $\left(\begin{smallmatrix} -1 & -2 & -3 \end{smallmatrix} \right)$ \\ \hline
\underline{11}  & $\mathfrak{su}(8)_{ps}$ & {\tiny $\left(
\begin{array}{ccccccc}
 0 & 1 & 0 & 0 & 1 & 0 & 0 \\
 1 & 0 & 0 & 1 & 0 & 0 & 0 \\
 0 & 0 & 0 & 0 & 0 & 0 & 0 \\
 -1 & -2 & -3 & -4 & -5 & -6 & -3 \\
\end{array}
\right)$}&{$\mathfrak{su}(2)_{\text{q}}\oplus \mathfrak{su}(2)_{\text{l}}\oplus\mathfrak{u}(1)_{B-L}$ } &{\tiny $\left(
\begin{array}{ccccccc}
1 & 2 & 3 & 2 & 1 & 0 & 0 \\
0 & 0 & 0 & 0 &0 & 0 & 1\\
 -1 & -2 & -3 & -4 & -5 & -6 & -3 \\
\end{array}
\right)$} \\ \hline
\underline{12}  & $\mathfrak{su}(12)_{ps}$ &{\tiny $\left(
\begin{array}{ccccccccccc}
 0 & 1 & 0 & 0 & 1 & 0 & 0 & 1 & 0 & 0 & 0 \\
 1 & 0 & 0 & 1 & 0 & 0 & 1 & 0 & 0 & 0 & 0 \\
 0 & 0 & 0 & 0 & 0 & 0 & 0 & 0 & 0 & 0 & 0 \\
 -1 & -2 & -3 & -4 & -5 & -6 & -7 & -8 & -9 & -6 & -3 \\
\end{array}
\right)$ }& { $\mathfrak{su}(3)_{\text{q}}\oplus \mathfrak{su}(3)_{\text{l}}\oplus \mathfrak{u}(1)_{B-L}$ }&{\tiny
$\left( \begin{array}{ccccccccccc}
0& 0 & 0 & 1 & 2 & 3 & 2 & 1 & 0 & 0  & 0\\
1 & 2 & 3 & 2 & 1 & 0 & 0 & 0 & 0 & 0 & 0\\
0 & 0 & 0& 0& 0 & 0 &0 &0 &0 & 0 & 1\\
0 & 0 & 0& 0& 0 & 0 &0 &0 &0 & 1 & 0\\
 -1 & -2 & -3 & -4 & -5 & -6 & -7 & -8 & -9 & -6 & -3 \\
\end{array}\right)$}\\ \hline
\end{tabular}
\end{adjustwidth}
\end{small}
\end{center}
\caption{ A set of building blocks $\{\mathfrak{b}_{\underline{a}}\}$, where $\underline{a}\in\{1,\dots,12\}$, from which all 27 maximal gauge algebras in Table~\protect\ref{tab:maximal} can be constructed.
For each building block $\mathfrak{b}_{\ua}$, we record the projection matrix $[\,\Lambda \alpha_{\underline{a}}\,]$ for the map from SM into $\mathfrak{b}_{\ua}$, as well as its centraliser $\mathfrak{C}_{\ua}$, and the projection matrix $[\,\Lambda \rho_{\underline{a}}\,]$ for the embedding of the centraliser in $\mathfrak{b}_{\ua}$.
\label{tab:blocks}}
\end{table}

For each of the projection matrices $\Lambda \alpha_{\underline{a}}$, we can find associated embeddings $\alpha_{\underline{a}}$ in the following way.
First, let us define the simple roots of $\sm$: let $h_Y$ be the generator of $u(1)_Y$, let $\kappa$ be the simple root of $\mathfrak{su}(2)$, and let $\{\tau_1, \tau_2\}$ be the simple roots of $\mathfrak{su}(3)$. Let $\Delta(\mathfrak{b}_{\ua}):=\{\lambda_r\}$ be the simple roots of $\mathfrak{b}_{\ua}$. 
For each Lie algebra we will work with the Chevalley basis. 
Roughly, given any semi-simple Lie algebra $\g$ we choose a basis $\{h_{\lambda_r}\}$ for its Cartan subalgebra that is dual to the fundamental weights, which when adjoined with a set of vectors $\{ e_\lambda\}$, where $\lambda$ is in the root system $\Phi(\g)$ of $\g$, forms a full basis of $\g$. For the details of this construction, and examples of Chevalley bases, are given in Appendix \ref{app:chevalley}.

A projection matrix $\Lambda \alpha_{\ua}$ tells us explicitly how its associated embedding $\alpha_{\ua}$ acts on the Cartan subalgebra of $\sm$. Namely, we have 
\begin{align}
\label{eq:alphaTau1}\alpha_{\ua}(h_{\tau_1})&=[\,\Lambda \alpha_{\ua}\,]_{1r}\,  h_{\lambda_r}\, ,\\
\alpha_{\ua}(h_{\tau_2})& =[\,\Lambda \alpha_{\ua}\,]_{2r}\,  h_{\lambda_r}\, ,\\
\alpha_{\ua}(h_{\kappa})&=[\,\Lambda \alpha_{\ua}\,]_{3r}\,  h_{\lambda_r}\, ,\\
\label{eq:alphaY}\alpha_{\ua}(h_Y)&=[\,\Lambda \alpha_{\ua}\,]_{4r}\,  h_{\lambda_r}\, ,
\end{align}
assuming summation on the index $r$.
The first two rows of the projection matrix tell us how $\su(3)$ is embedded, the third row tells us how $\su(2)_L$ is embedded, and the last row tells us how hypercharge is embedded.

To find out how the embedding $\alpha_{\ua}$ acts on the non-Cartan basis elements, we define, for each $\tau \in \Phi(\su(3))$ and for $\pm \kappa$ (the roots of $\su(2)$), the following sets {(see \emph{e.g.}~\cite{GRUBER197595})}
\begin{align}
\Gamma_\tau&:=\{ \lambda \in \Phi(\g)\mid (\Lambda \alpha_{\ua}\,|_{\su(3)})\lambda=\tau\}\, ,\\
\Gamma_{\pm \kappa}&:=\{ \lambda \in \Phi(\g)\mid (\Lambda \alpha_{\ua}\,|_{\su(2)})\lambda=\pm \kappa\}\, .
\end{align}
For us, the intersection of any two such $\Gamma_\tau$ or $\Gamma_{\pm \kappa}$ is always the empty set. The image of $\alpha_{\ua}$ is then given by 
\begin{align}
\alpha_{\ua}(e_\tau)=\sum_{\lambda\in \Gamma_\tau} f(\lambda) e_\lambda,\\
\alpha_{\ua}(e_{\pm \kappa})=\sum_{\lambda \in \Gamma_{\pm \kappa}} f(\lambda) e_\lambda
\end{align}
for some map of sets $f: \Phi(\b_{\ua}) \rightarrow \C$. Conditions are put on $f$ by ensuring that $\alpha_{\ua}$ preserves commutators. Any such $f$ satisfying these conditions provides a valid choice of $\alpha_{\ua}$. Different valid choices are related and will lead to the same final result (a consequence of the theorems proven in Appendix B of Ref.~\cite{Tooby-Smith:2021hjk}).

\subsubsection{Example: $\mathfrak{so}(6)_R$}\label{sec:so6I}

We now give a couple of explicit examples to illustrate how the embeddings $\alpha_{\ua}$ for the various building blocks $\mathfrak{b}_{\ua}$ listed in Table \ref{tab:blocks} can be inferred from the projection matrices $\Lambda \alpha_{\ua}$. For our first example, which is especially straightforward, consider the building block $\b_{\underline{6}}=\so(6)_R$.
As can be read from Table~\ref{tab:blocks}, the projection matrix for $\mathfrak{so}(6)_R$ is given by 
\begin{align}
[\,\Lambda\alpha_{\underline{6}}\,]=\left(\begin{smallmatrix} 0&0&0\\0&0&0\\0&0&0\\ \frac{9}{2}& 3& \frac{3}{2}\end{smallmatrix}\right)\, .
\end{align}
From this, it is easy to read off $\alpha_{\underline{6}}$ since it acts trivially on $\mathfrak{su}(2)_L$ and $\mathfrak{su}(3)$, whilst mapping the hypercharge generator to
\begin{align}
\alpha_{\underline{6}}(h_{Y})=\frac{9}{2}h_{\lambda_1} + 3h_{\lambda_2}+ \frac{3}{2}h_{\lambda_3}\, .
\end{align}

\subsubsection{Example: $\mathfrak{sp}(4)_L$}

For a second, slightly more involved, example, consider the building block $\b_{\underline{8}}=\sp(4)_L$, that appears in models with electroweak flavour unification~\cite{Davighi:2022fer} in two generations.
The projection matrix for the embedding $\alpha_{\underline{8}}:\mathfrak{sm}\rightarrow \mathfrak{sp}(4)_L$ 
is
\begin{align} \label{eq:LambdaAlpha8}
[\,\Lambda\alpha_{\underline{8}}\,]=\left(\begin{smallmatrix}0 & 0 \\ 0 &0 \\ 1&2 \\ 0&0\end{smallmatrix}\right).
\end{align}
From this we see that only $\mathfrak{su}(2)_L$ embeds non-trivially into $\mathfrak{sp}(4)_L$, and we can ignore the other summands in $\sm$. Eq. (\ref{eq:LambdaAlpha8}) tells us that 
\begin{align}
\alpha_{\underline{8}}(h_{\kappa})&=h_{\lambda_1}+2h_{\lambda_2}\,, \\
\alpha_{\underline{8}}(e_{\kappa})&=f(\lambda_2) e_{\lambda_2} + f(\lambda_1+\lambda_2) e_{\lambda_1+\lambda_2} + f(2\lambda_1+\lambda_2) e_{2\lambda_1+\lambda_2}\,,\\
\alpha_{\underline{8}}(e_{-\kappa})&=f(-\lambda_2) e_{-\lambda_2} + f(-\lambda_1-\lambda_2) e_{-\lambda_1-\lambda_2} + f(-2\lambda_1-\lambda_2) e_{-2\lambda_1-\lambda_2}\,.
\end{align}
From the fact that $\alpha_{\underline{8}}$ must be an embedding, the map $f$ must satisfy one of two sets of conditions. Either
\begin{align}
f(\lambda_{1}+\lambda_2)f(-\lambda_{1}-\lambda_2)\ne 0, \quad f(2\lambda_1+\lambda_2) f(-\lambda_1-\lambda_2)=-f(-\lambda_2) f(\lambda_1+\lambda_2), \nonumber \\
f(-2\lambda_1-\lambda_2) f(\lambda_1+\lambda_2)=-f(\lambda_2)f(-\lambda_1-\lambda_2)\, , \nonumber\\
f(-\lambda_1-\lambda_2)f(\lambda_1+\lambda_2)+ f(\nu_2) f(-\nu_2)=1\, ,
\end{align}
or 
\begin{align}
f(\pm (\lambda_1+\lambda_2))=0,\quad f(\lambda_2) f(-\lambda_2)=1,\quad f(2\lambda_1+\lambda_2)f(-2 \lambda_1-\lambda_2)=1.
\end{align}
We choose the values 
\begin{align}
f(\pm(2\lambda_1+\lambda_2))=f(\pm \lambda_2)=0\, , \qquad f(\pm (\lambda_1+\lambda_2))=1.
\end{align}
This gives us the valid embedding of $\su(2)_L$ in $\sp(4)_L$:
\begin{align}
\alpha_{\underline{8}}(h_\kappa)&=h_{\lambda_1}+2h_{\lambda_2}\\
\alpha_{\underline{8}}(e_\kappa)&=e_{\lambda_1+\lambda_2}\\
\alpha_{\underline{8}}(e_{-\kappa})&=e_{-\lambda_1-\lambda_2}.
\end{align}
Using similar methods, one can find valid embeddings $\alpha_{\ua}$ for all 12 of the building blocks $\mathfrak{b}_{\ua}$ listed in Table~\ref{tab:blocks}. 

\subsubsection{Assembling the building blocks (I)}

From there, it is easy to construct the embeddings $\alpha_i:\sm \to \g_i$ for each of the 27 maximal algebras $\g_i$ in Table~\ref{tab:maximal}.
In particular, the corresponding projection matrix is given by
\begin{align} \label{eq:assemble_embedding}
[\,\Lambda \alpha_i\,]= \begin{pmatrix} [\,\Lambda\alpha_{\underline{a_1}}\,] & [\,\Lambda\alpha_{\underline{a_2}}\,]  & [\,\Lambda\alpha_{\underline{a_3}} \,] & \cdots & [\,\Lambda\alpha_{\underline{a_n}} \,] & [\, {\bf 0}_{4\times \mathrm{rank}(\mathfrak{hor}_i)}\, ]\end{pmatrix}\, ,
\end{align}
where $I_i=\{\underline{a_1}, \underline{a_2}, \underline{a_3}, \dots, \underline{a_n}\}$.

\subsubsection{Example: $\g_{27}=\su(5)^{\oplus 2} \oplus \su(2)^{\oplus 2}$} \label{sec:g27_1}
For the algebra $\g_{27}$, recall we have $I_{27}=\{\ul{1},\ul{1}\}$, and $\hor_{27}=\su(2)^{\oplus 2}$. The projection matrix $[\Lambda \alpha_{27}]$ can therefore be got from Eq.~(\ref{eq:assemble_embedding}) and Table.~\ref{tab:maximal}. Explicitly,
\begin{align}\label{eq:alpha27}
[\Lambda \alpha_{27}]=\left(
\begin{array}{cccc|cccc|cc}
 0 & 0 & 1 & 0 & 0 & 0 & 1 & 0 & 0 & 0 \\
 0 & 0 & 0 & 1 & 0 & 0 & 0 & 1 & 0 & 0 \\
 1 & 0 & 0 & 0 & 1 & 0 & 0 & 0 & 0 & 0 \\
 3 & 6 & 4 & 2 & 3 & 6 & 4 & 2 & 0 & 0 \\
\end{array}
\right)\, .
\end{align}
We do not need $\alpha_{27}$ in what follows explicitly so will not report it here.
\subsection{Step 2: Computing the centralisers of $\sm$ in $\g$} \label{sec:centraliser}

Given an embedding $\alpha_i:\sm \to \g_i$, the next step is to compute the centraliser of $\sm$ in $\g_i$. 
Fixing $\sm$, for each maximal gauge algebra $\g_i$ the centraliser is itself a pair $(\mathfrak{C}_{i},\rho_i)$ consisting of an algebra $\mathfrak{C}_i:=\mathfrak{C}_{\g_i}(\sm)$, as defined in Eq.~(\ref{eq:Cg_def}), and another embedding $\rho_i:\mathfrak{C}_i\rightarrow \g_i$. 

The factorisation of $\g_i$ into the building blocks $\{\mathfrak{b}_{\ua}\}$, as in (\ref{eq:gi_decomp}), implies there is a similar factorisation of the centraliser $(\mathfrak{C}_i,\rho_i)$
\begin{align} \label{eq:build_C}
\mathfrak{C}_i &= \bigoplus_{\underline{a}\in I_i} \mathfrak{C}_{\ua} \oplus \mathfrak{hor}_i \, , \\
\rho_i &= \bigoplus_{\underline{a}\in I_i} \rho_{\ua} \oplus \rho_{\mathfrak{hor}_i} \, ,
\end{align}
where $(\mathfrak{C}_{\ua},\rho_{\ua})$ is the centraliser of $\alpha_{\ua}$.
Clearly, the centraliser of $\mathfrak{hor}_i$ is isomorphic to itself. Thus, the embedding $\rho_{\mathfrak{hor}_i}$ can be chosen to be the identity.

Given our knowledge of all the embeddings $\{\alpha_{\ua}\}$, the task of finding each `building block centraliser' $\mathfrak{C}_{\ua}$ becomes reasonably straightforward. It simply amounts to finding all elements of $\mathfrak{b}_{\ua}$ which commute with the image of $\alpha_{\ua}$. 
Regarding the embedding $\rho_{\ua}$, all the information of interest to us is captured by the projection matrix $[\,\Lambda \rho_{\ua}\,]$. 
The centralisers $\mathfrak{C}_{\ua}$ and associated $[\,\Lambda \rho_{\ua}\,]$ for each of our building blocks $\mathfrak{b}_{\ua}$ are displayed in Table~\ref{tab:blocks}. In the next two Subsections we show how these are calculated, continuing the explicit examples of $\mathfrak{b}_{\underline{6}}=\so(6)_R$ and $\mathfrak{b}_{\underline{8}}=\sp(4)_L$. We continue to use the Chevalley basis 
for each Lie algebra.

\subsubsection{Example (contd.): $\mathfrak{so}(6)_R$} \label{sec:so6II}
A general element of the Lie algebra $\mathfrak{b}_{\underline{6}}=\mathfrak{so}(6)_R$, which is 15-dimensional, is expanded in the Chevalley basis as follows,
\begin{align}
\mathfrak{so}(6)_R \ni u=\, &c_1 h_{\lambda_1} +c_2 h_{\lambda_2} +c_3 h_{\lambda_3} +c_4 e_{\lambda_{123}} +c_5 e_{\lambda_{23}} + c_6 e_{\lambda_{12}} +c_7 e_{\lambda_2} + c_8 e_{\lambda_3} + c_9 e_{\lambda_1} \nonumber \\
&+c_{10} e_{-\lambda_{123}} +c_{11} e_{-\lambda_{23}} + c_{12} e_{-\lambda_{12}} +c_{13} e_{-\lambda_2} + c_{14} e_{-\lambda_3} + c_{15} e_{-\lambda_1}\, ,
\end{align}
where each coefficient $c_i\in \C${, and \emph{e.g.} $\lambda_{123}:=\lambda_1+\lambda_2+\lambda_3$}. We want to find all such $u$ that satisfy the condition 
\begin{align}
[u,\alpha_{\underline{6}}(h_Y)]=0.
\end{align}
The most general $u$ which satisfies this condition is  given by 
\begin{align}
u=c_1 h_{\lambda_1} +c_2 h_{\lambda_2} +c_3 h_{\lambda_3} +c_5 e_{\lambda_{12}}+c_7 e_{\lambda_2} + c_8 e_{\lambda_3} +c_{11} e_{-\lambda_{12}} + c_{13} e_{-\lambda_2}+ c_{14} c_{-\lambda_3}.
\end{align} 
The space of such $u$ defines a Lie subalgebra of $\so(6)_R$, the centraliser of $\u_Y$, which can be shown to be isomorphic to $\mathfrak{C}_{\underline{6}}= \mathfrak{su}(3)\oplus \mathfrak{u}(1)$. 
The $\su(3)$ here is a family symmetry that acts by complex rotations on the family index of each right-handed species of field, {\em i.e.} on {$d_i$, $u_i$, $e_i$, and $\nu_i$.} (But note that this information, namely how $\mathfrak{C}_{\underline{6}}$ acts on the fermions, cannot yet be deduced; doing so requires the steps detailed in Section~\ref{sec:flatland}.)
The $\u$ summand in $\mathfrak{C}_{\underline{6}}$ is of course just $\u_Y$.

The embedding $\rho_{\underline{6}}:\mathfrak{C}_{\underline{6}}\rightarrow \so(6)_R$ is given by 
\begin{align}
&\rho_{\underline{6}}(h_{\tau_1})=h_{\lambda_2}\,,
&&\rho_{\underline{6}}(h_{\tau_2})=h_{\lambda_3}\,,\\
&\rho_{\underline{6}}(e_{\pm\tau_1})=e_{\pm\lambda_2}\,, &&\rho_{\underline{6}}(e_{\pm\tau_2})=e_{\pm\lambda_3}\,, \\
&\rho_{\underline{6}}(e_{\pm\tau_{12}})=-e_{\pm \lambda_{23}}
\end{align}
for the $\su(3)$ part of $\mathfrak{C}_{\underline{6}}$, where the $\su(3)$ generators are given in the Chevalley basis (which the reader can find in examples given in Appendix~\ref{app:chevalley_su4}), and of course
\begin{align}
\rho_{\underline{6}}(h_Y)&=\alpha_{\underline{6}}(h_Y)=\frac{9}{2} h_{\lambda_1} + 3h_{\lambda_2} + \frac{3}{2} h_{\lambda_3}
\end{align} 
for $\u_Y$.
From this we can read off the projection matrix $[\Lambda \rho_{\underline{6}}]$ to be 
\begin{align}
[\,\Lambda \rho_{\underline{6}}\,]= \begin{pmatrix} 0 & 1 & 0 \\ 0 & 0 & 1 \\ \frac{9}{2} & 3 & \frac{3}{2} \end{pmatrix}.
\end{align}

\subsubsection{Example (contd.): $\mathfrak{sp}(4)_L$}
Turning to our second example, a general element of the Lie algebra $\mathfrak{b}_{\underline{8}}=\mathfrak{sp}(4)_L$ is given by 
\begin{align}
\mathfrak{sp}(4)_L \ni
v=\, &c_1h_{\lambda_1} + c_2h_{\lambda_2} +c_3 e_{2\lambda_1+\lambda_2} + c_4e_{\lambda_1+\lambda_2} +c_5 e_{\lambda_2} +c_6 e_{\lambda_1} \nonumber \\
&+ c_7e_{-2\lambda_1-\lambda_2} +c_8 e_{-\lambda_1-\lambda_2}+c_9 e_{-\lambda_2} +c_{10} e_{-\lambda_1}
\end{align}
Here we want to enforce the condition that 
\begin{align}
[ v,\alpha_{\underline{8}}(h_{\kappa})]=0, \quad [ v, \alpha_{\underline{8}}(e_{\kappa})]=0, \quad [ v, \alpha_{\underline{8}}(e_{-\kappa})]=0,
\end{align}
{\em i.e.} that $v$ commutes with the image of $\su(2)_L$ in $\sp(4)_L$.
The most general $v$ which satisfies all these conditions is simply 
\begin{align}
v=c_1h_{\lambda_1}\, .
\end{align}
Thus, we have that the centraliser is a 1-parameter subalgebra of $\sp(4)_L$, $\mathfrak{C}_{\underline{8}}=\mathfrak{u}(1)$. Letting $h_X$ denote its generator, we have
\begin{align}
\rho_{\underline{8}}(h_X)=h_{\lambda_1}
\end{align}
which tells us that 
\begin{align}
[\Lambda \rho_{\underline{8}}]=\begin{pmatrix} 1& 0 \end{pmatrix}.
\end{align}
{Physically, this $\u$ acts as a rotation in family-space on (two families of) the {$q$} and {$\ell$} left-handed $SU(2)_L$ doublets.}

\subsubsection{Assembling the building blocks (II)}

Once we have computed $\Lambda \rho_{\underline{a}}$ for each building block $\mathfrak{b}_{\ua}$ (as per the two examples above), one can then build the projection operator 
\begin{align} \label{eq:build_rho}
\Lambda \rho_i=\bigoplus_{\underline{a}\in I_i} \Lambda \rho_{\underline{a}} \oplus \Lambda\rho_{\mathfrak{hor}_i}
\end{align}
associated to each $\g_i$. For the horizontal factor, the projection operator $\rho_{\mathfrak{hor}_i}$ is simply the identity map.
In terms of matrices (in our chosen basis), 
\begin{align} \label{eq:assemble_rho}
[\,\Lambda \rho_i\,]= 
\begin{pmatrix} 
[\,\Lambda\rho_{\underline{a_1}}\,] & 0 & 0 & \cdots & 0 &0\\
0 & [\,\Lambda\rho_{\underline{a_2}}\,] & 0 & \cdots & 0 &0\\
0 & 0 & [\,\Lambda\rho_{\underline{a_3}} \,] & \cdots & 0 &0\\
\cdots & \cdots & \cdots & \cdots & \cdots & \cdots \\
0 & 0 & 0 & \cdots & [\,\Lambda\rho_{\underline{a_n}}\,] & 0\\
0 & 0 & 0 & \cdots & 0 & [\, {\bf 1}_{\mathrm{rank}(\mathfrak{hor}_i)}\, ]
\end{pmatrix}\, ,
\end{align}
where again $I_i=\{\underline{a_1}, \underline{a_2}, \underline{a_3}, \dots, \underline{a_n}\}$, and where ${\bf 1}_{\mathrm{rank}(\mathfrak{hor}_i)}$ denotes the $\mathrm{rank}(\mathfrak{hor}_i) \times \mathrm{rank}(\mathfrak{hor}_i)$ identity matrix. Note that the projection matrix here is block diagonal, which should be contrasted with the `horizontal concatenation' used in the corresponding Eq.~(\ref{eq:assemble_embedding}) that assembles the projection matrix for the embedding.\footnote{Lest there is any confusion, the reason for the difference is that the projections $[\Lambda \alpha_{\ua}]$ for each embedding $\alpha_{\ua}:\sm \to \mathfrak{b}_{\ua}$ has a row index that always runs over the four Cartan generators of $\sm$ (while the column index runs over the Cartan of $\mathfrak{b}_{\ua}$), and so the number of rows is unchanged upon concatenating the building blocks. Here, on the other hand, each $\rho_{\ua}$ is a map from a different summand $C_{\ua}$ of the centraliser $C_i$, to the corresponding $\mathfrak{b}_{\ua}$; thus, to assemble the building blocks, the projection is here a direct sum.
 }
\subsubsection{Example (contd. II): $\g_{27}=\su(5)^{\oplus 2} \oplus \su(2)^{\oplus 2}$} \label{sec:g27_2}
Recall that for $\g_{27}$ we have $I_{27}=\{\ul{1},\ul{1}\}$, and $\hor_{27}=\su(2)^{\oplus 2}$. 
Using the results for the building blocks (Table~\ref{tab:blocks}), the centraliser is
\begin{align}
\mathfrak{C}_{27} = \u^{\oplus 2} \oplus \su(2)^{\oplus 2}\, .
\end{align} 
Then from Eq.~(\ref{eq:assemble_rho}) and Table~\ref{tab:blocks} we have that 
\begin{align}\label{eq:rho27}
[\Lambda \rho_{27}]=\left(
\begin{array}{cccc|cccc|cc}
 3 & 6 & 4 & 2 & 0 & 0 & 0 & 0 & 0 & 0 \\ \hline
 0 & 0 & 0 & 0 & 3 & 6 & 4 & 2 & 0 & 0 \\ \hline
 0 & 0 & 0 & 0 & 0 & 0 & 0 & 0 & 1 & 0 \\ 
 0 & 0 & 0 & 0 & 0 & 0 & 0 & 0 & 0 & 1 \\
\end{array}
\right)\, .
\end{align}

\subsection{Step 3: Finding the planes of charges} \label{sec:flatland}

Let us briefly summarise what we have achieved so far: in Section~\ref{sec:embeddings} we showed how to construct the embeddings $\alpha_i:\sm \to \g_i$ for each of the 27 maximal algebras in Table~\ref{tab:maximal}; then in Section~\ref{sec:centraliser} we showed how, given these embeddings, to compute the centralisers $(\mathfrak{C}_i,\rho_i)$ of the $\sm$ in each $\g_i$. 
The remaining step, which is the subject of this Subsection, is to parametrise the plane of charges associated with a Cartan $\h_{\mathfrak{C}_i}$. To do this, we also need the information encoded in the maps $\beta_i:\g_i \to \su(48)$, defined in Eq.~(\ref{eq:triangle}), which specify how the SM fermions are embedded in each $\g_i$.

In~\cite{Allanach:2021bfe}, the maps $\beta_i:\g_i \to \su(48)$ are not given explicitly. Rather, the program in Ref.~\cite{Allanach:2021bfe} outputs the {\em representation} of $\g_i$ that each $\beta_i$ corresponds to. From this information, we can find the weight system of the representation, which we denote $\Phi(\beta_i)$.  Since the weights tell us the eigenvalues of all fermion components in the representation under the Cartan subalgebra $\h_{\g_i}$, the weight system contains the information we need to extract the charges of each fermion under $\h_{\mathfrak{C}_i}$.

To do this, let us first define an indexing set of SM fermion fields,
\begin{align}\label{eq:defQ}
\mathcal{Q}:=\{F_j \mid F\in \{q,u^c,d^c,\ell,e^c,\nu^c\} \text{ and }j\in \{1,2,3\}\}.
\end{align}
Let us also define the weights $\tilde w_F\in \Phi(\gamma)$ for $F\in \{q,u^c,d^c,\ell,e^c,\nu^c\}$ as the highest weights of the SM fermionic representations, where recall $\gamma:\sm \to \su(48)$ as defined below Eq.~(\ref{eq:triangle}). Namely, 
\begin{align}
[\tilde w_q]=\begin{pmatrix} 1 & 0 & 1 & 1 \end{pmatrix}, \quad [\tilde w_{u^c}]=\begin{pmatrix} 0& 1 & 0 & -4 \end{pmatrix},\quad [\tilde w_{d^c}]=\begin{pmatrix} 0 & 1 & 0 & 2 \end{pmatrix},\\
[\tilde w_{\ell}]=\begin{pmatrix} 0 & 0 & 1 & -3 \end{pmatrix},\quad [\tilde w_{e^c}]=\begin{pmatrix} 0 & 0 & 0 & 6 \end{pmatrix},\quad [\tilde w_{\nu^c}]=\begin{pmatrix} 0 & 0 & 0 & 0 \end{pmatrix}.
\end{align}
Then for each $F_j\in \mathcal{Q}$ we can assign a choice of weight $w_{F_j} \in \Phi(\beta_i)$ such that no two $w_{F_j}$'s are the same, and 
\begin{align}\label{eq:wpsiCond}
w_{F_j}\circ \alpha_i=\tilde w_{F}\, .
\end{align}
This condition can be checked using the projection matrices, since it holds if and only if 
\begin{align}
[\Lambda \alpha_i][w_{F_j}]=[\tilde w_F]\, . 
\end{align}
There is of course freedom to permute the family labels here, for example one could assign a given weight to $w_{q_1}$ or $w_{q_2}$. For each $w_\psi$ there are exactly three possible choices in $\Phi(\beta_i)$, so each possible choice appears exactly once.

We define a vector space $\Q^{18}$ with a basis $\{b_\psi\}_{\psi \in \mathcal{Q}}$. This vector space can be interpreted as our `space of charges'. The $\sm\oplus \u_X$ anomaly cancellation conditions (ACCs), for example, define a complicated surface in this space, whilst here we will construct planes within it. There is a linear map $\chi$ from $\h_{\g_i}^\mathbb{Q}$, which we define to be the rational space spanned by $\{h_{\lambda}\}_{\lambda\in \Delta(\g_i)}$, to $\mathbb{Q}^{18}$, defined by our choice of $w_\psi$. Namely, letting $h\in \h_{\g_i}^\Q$, the map $\chi:\h_{\g_i}^\Q\rightarrow \Q^{18}$ is defined through
\begin{align}
\chi(h)=\sum_{\psi\in \mathcal{Q}} w_\psi(h)b_\psi.
\end{align}
The physics interpretation of this equation is that the `component' $w_\psi(h)$ in the $b_\psi$ direction is the `charge' of the fermion $\psi$ with respect to any Cartan generator $h\in  \h^\Q_{\g_i}$ (where, at the moment, this includes the generators corresponding to the embedding of $\sm$).

The embedding of the centraliser $\rho_i:\mathfrak{C}_i \to \g_i$,  then defines a subplane of $\mathbb{Q}^{18}$, which is given by
\begin{align}
P_i:=\chi\left(\h_{\g_i}^\mathbb{Q}\cap \mathrm{Im}(\rho_i)\right).
\end{align}
Each point in $P_i$ corresponds to the charges of all 18 fermions under a particular element in the Cartan of $\mathfrak{C}_i$.
In this way, we construct the planes $P_i$ that parametrise the fermion charges of any abelian extension $\epsilon:\sm_S \oplus \a \to \su(48)$ that embeds in $\g_i$, for each of the maximal $\g_i$ listed in Table~\ref{tab:maximal}. We emphasise that the generator of $\u_Y$ (that will remain unbroken) is included in this plane.

To be more explicit, it is easy to show that the plane $P_i$ is spanned by the $d_i:=\mathrm{rank} (\h_{\mathfrak{C}_i})$ vectors 
\begin{align} \label{eq:Xik}
S^{i,k}=\sum_{\psi\in \mathcal{Q}} ([\Lambda\rho_i]_{k\ast}\cdot [w_\psi]) b_\psi\, , \qquad k\in\{1,\dots, d_i\}\, ,
\end{align}
where $[\Lambda\rho_i]_{k\ast}$ denotes the $k$th row of the projection matrix $[\Lambda\rho_i]$ that is constructed as in Eq.~(\ref{eq:assemble_rho}) using the results in Table~\ref{tab:blocks}. One can interpret the vectors $S^{i,k}\in \Q^{18}$, for each value of $k$, as a basis for the possible  $u_{X^1}\oplus \u_{X^2} \oplus \dots$  charges which can be embedded into the $\g_i$ model. The dimension of $P_i$ is $d_i$.

\begin{table}
\begin{equation*}
\begin{array}{|c|ccc|ccc|ccc|ccc|ccc|ccc|}
\hline
& q_1 & q_2 & q_3 & u_1^c& u_2^c & u_3^c & d_1^c & d_2^c & d_3^c & \ell_1 & \ell_2 & \ell_3 &e_1^c & e_2^c &e_3^c & \nu_1^c &\nu_2^c &\nu_3^c\\ \hline
 Y& 1 & 1 & 1 & -4 & -4 & -4 & 2 & 2 & 2 & -3 & -3 & -3 & 6 & 6 & 6 & 0 & 0 & 0 \\
 Y_1 & 1 & 0 & 0 & -4 & 0 & 0 & 2 & 0 & 0 & -3 & 0 & 0 & 6 & 0 & 0 & 0 & 0 & 0\\
 Y_2 &0 & 1 & 0 & 0 & -4 & 0 & 0 & 2 & 0 & 0 & -3 & 0 & 0 & 6 & 0 & 0 & 0 & 0\\ 
 Y_3 & 0 & 0 & 1 & 0 & 0 & -4 & 0 & 0 & 2 & 0 & 0 & -3 & 0 & 0 & 6 & 0 & 0 & 0 \\
 \hline
BL& 1 & 1 & 1 & -1 & -1 & -1 & -1 & -1 & -1 & -3 & -3 & -3 & 3 & 3 & 3 & 3 & 3 & 3 \\
 BL_3 & 0 & 0 & 1 & 0 & 0 & -1 & 0 & 0 & -1 & 0 & 0 & -3 & 0 & 0 & 3 & 0 & 0 & 3 \\
 \hline
  B_{12}& 1 & -1 & 0 & -1 & 1 & 0 & -1 & 1 & 0 & 0 & 0 & 0 & 0 & 0 & 0 & 0 & 0 & 0 \\
 B_{23}& 0 & 1 & -1 & 0 & -1 & 1 & 0 & -1 & 1 & 0 & 0 & 0 & 0 & 0 & 0 & 0 & 0 & 0 \\

 \hline
 L_{12} & 0 & 0 & 0 & 0 & 0 & 0 & 0 & 0 & 0 & 1 & -1 & 0 & -1 & 1 & 0 & -1 & 1 & 0 \\
 L_{23} & 0 & 0 & 0 & 0 & 0 & 0 & 0 & 0 & 0 & 0 & 1 & -1 & 0 & -1 & 1 & 0 & -1 & 1 \\
 \hline
 S_{12} & 1 & -1 & 0 & 0 & 0 & 0 & 0 & 0 & 0 & 1 & -1 & 0 & 0 & 0 & 0 & 0 & 0 & 0\\
D_{12} & 0 & 0 & 0 & 1 & -1 & 0 & -1 & 1 & 0 & 0 & 0 & 0 & -1 & 1 & 0 & 1 & -1 & 0 \\
D_{23} & 0 & 0 & 0 & 0 & 1 & -1 & 0 & -1 & 1 & 0 & 0 & 0 & 0 & -1 & 1 & 0 & 1 & -1 \\

\hline
 F_{12} & 0 & 0 & 0 & 0 & 0 & 0 & 1 & -1 & 0 & 1 & -1 & 0 & 0 & 0 & 0 & 0 & 0 & 0 \\
 T_{12} &1 & -1 & 0 & 1 & -1 & 0 & 0 & 0 & 0 & 0 & 0 & 0 & 1 & -1 & 0 & 0 & 0 & 0 \\
 N_{12} & 0 & 0 & 0 & 0 & 0 & 0 & 0 & 0 & 0 & 0 & 0 & 0 & 0 & 0 & 0 & 1 & -1 & 0 \\
 \hline
\end{array}
\end{equation*}
\caption{Charges of the SM$+3\nu_R$ fermions under a set of $\u_X$ symmetries that play a central role in this paper. Namely, every possible $\sm\oplus \u_X$ gauge model that embeds inside a semi-simple $\g$ can be expressed as a linear combination of a particular subset of these rows (that we determine), up to family permutations within each species. Here `$B$' and `$L$' stand for \emph{baryon} and \emph{lepton} as usual, `$Y$' stands for (global) \emph{hypercharge}, `$S$' stands for \emph{sinistral} (meaning left-handed), `$D$' stands for \emph{dextral} (meaning right-handed), `$F$' stands for \emph{five}, `$T$' for \emph{ten}, and `$N$' for \emph{neutrino}. \label{tab:charges}
}
\end{table}

\subsubsection{Example (contd. III): $\g_{27}=\su(5)^{\oplus 2} \oplus \su(2)^{\oplus 2}$} \label{sec:g27_3}

We now illustrate this construction by continuing with the explicit example of $\g_{27}$.
The representation corresponding to $\beta_{27}$, which can be read off from Table~\ref{tab:maximal}, is
\begin{align}
\beta_{27}\sim (\overline{\mathbf{5}}, {\mathbf{1}}, {\mathbf{1}}, {\mathbf{1}})^{\oplus 2}  \oplus ({\mathbf{1}}, \overline{\mathbf{5}}, {\mathbf{1}}, {\mathbf{1}}) \oplus ({\mathbf{1}}, {\mathbf{10}}, {\mathbf{1}}, {\mathbf{1}}) \oplus ({\mathbf{10}}, {\mathbf{1}}, {\mathbf{2}}, {\mathbf{1}}) \oplus ({\mathbf{1}}, {\mathbf{1}}, {\mathbf{1}}, {\mathbf{3}}).
\end{align}
From this, it is possible to find the weight system $\Phi(\beta_{27})$ which contains $48$ weights, and as such we will not report it in full here.

Using $[\Lambda\alpha_{27}]$ in Eq.~(\ref{eq:alpha27}) we can make a choice of weights $w_\psi\in \Phi(\beta_{27})$ which satisfy our condition in Eq.~(\ref{eq:wpsiCond}). A valid such choice is
\begin{align} \label{eq:g27_weights}
\begin{array}{c}
{[}w_{q_1}{]}=(\\
{[}w_{q_2}{]}=(\\
{[}w_{q_3}{]}=(\\
{[}w_{u^c_1}{]}=(\\
{[}w_{u^c_2}{]}=(\\
{[}w_{u^c_3}{]}=(\\
{[}w_{d^c_1}{]}=(\\
{[}w_{d^c_2}{]}=(\\
{[}w_{d^c_3}{]}=(
\end{array}
\begin{array}{ccccccccccc}
 0 & 0 & 0 & 0 & 1 & -1 & 1 & 0 & 0 & 0 \\
 1 & -1 & 1 & 0 & 0 & 0 & 0 & 0 & -1 & 0 \\
 1 & -1 & 1 & 0 & 0 & 0 & 0 & 0 & 1 & 0 \\
 0 & 0 & 0 & 0 & 0 & -1 & 0 & 1 & 0 & 0 \\
 0 & -1 & 0 & 1 & 0 & 0 & 0 & 0 & 1 & 0 \\
 0 & -1 & 0 & 1 & 0 & 0 & 0 & 0 & -1 & 0 \\
 0 & 0 & 0 & 1 & 0 & 0 & 0 & 0 & 0 & 0 \\
 0 & 0 & 0 & 1 & 0 & 0 & 0 & 0 & 0 & 0 \\
 0 & 0 & 0 & 0 & 0 & 0 & 0 & 1 & 0 & 0 \\
\end{array}
\begin{array}{c}
),\\
),\\
),\\
),\\
),\\
),\\
),\\
),\\
),\\
\end{array}
\quad
\begin{array}{c}
{[}w_{\ell_1}{]}=(\\
{[}w_{\ell_2}{]}=(\\
{[}w_{\ell_3}{]}=(\\
{[}w_{e^c_1}{]}=(\\
{[}w_{e^c_2}{]}=(\\
{[}w_{e^c_3}{]}=(\\
{[}w_{\nu^c_1}{]}=(\\
{[}w_{\nu^c_2}{]}=(\\
{[}w_{\nu^c_3}{]}=(
\end{array}
\begin{array}{ccccccccccc}
 1 & -1 & 0 & 0 & 0 & 0 & 0 & 0 & 0 & 0 \\
 1 & -1 & 0 & 0 & 0 & 0 & 0 & 0 & 0 & 0 \\
 0 & 0 & 0 & 0 & 1 & -1 & 0 & 0 & 0 & 0 \\
 0 & 0 & 0 & 0 & 0 & 1 & 0 & 0 & 0 & 0 \\
 0 & 1 & 0 & 0 & 0 & 0 & 0 & 0 & 1 & 0 \\
 0 & 1 & 0 & 0 & 0 & 0 & 0 & 0 & -1 & 0 \\
 0 & 0 & 0 & 0 & 0 & 0 & 0 & 0 & 0 & 2 \\
 0 & 0 & 0 & 0 & 0 & 0 & 0 & 0 & 0 & 0 \\
 0 & 0 & 0 & 0 & 0 & 0 & 0 & 0 & 0 & -2 \\
\end{array}
\begin{array}{c}
),\\
),\\
),\\
),\\
),\\
),\\
),\\
),\\
),
\end{array}
\end{align}
where each weight has 10 columns corresponding to the 10 fundamental weights of $\su(5)^{\oplus 2} \oplus \su(2)^{\oplus 2}$, using the same ordering as in Subsections~\ref{sec:g27_1} and~\ref{sec:g27_2}.
We have assigned the weights (\ref{eq:g27_weights}) such that $w_{F_1}< w_{F_2}< w_{F_3}$ lexicographically.\footnote{Lexicographically here means that $\{a_1,a_2, \cdots, a_n\}< \{b_1,b_2, \cdots, b_n\}$ if and only if there is an $i\in \{1,\ldots n\}$ such that $a_i<b_i$ and $a_j=b_j$ for $j<i$. It is the standard ordering of lists in many computer programs.}

Since the rank of the projection matrix $[\Lambda \rho_{27}]$ is $4$, the plane $P_{27}$ has dimension $d_{27}=4$. Using our projection matrix $[\Lambda \rho_{27}]$ in Eq.~(\ref{eq:rho27}) and the equations for $S^{i,k}$ in Eq.~(\ref{eq:Xik}), we compute the set of vectors
\begin{align}
\begin{array}{c}
S^{27,1}=(\\
S^{27,2}=(\\
S^{27,3}=(\\
S^{27,4}=(\\
\end{array}
\begin{array}{ccc|ccc|ccc|ccc|ccc|ccc}
 0 & 1 & 1 & 0 & -4 & -4 & 2 & 2 & 0 & -3 & -3 & 0 & 0 & 6 & 6 & 0 & 0 & 0 \\
 1 & 0 & 0 & -4 & 0 & 0 & 0 & 0 & 2 & 0 & 0 & -3 & 6 & 0 & 0 & 0 & 0 & 0 \\
 0 & -1 & 1 & 0 & 1 & -1 & 0 & 0 & 0 & 0 & 0 & 0 & 0 & 1 & -1 & 0 & 0 & 0 \\
 0 & 0 & 0 & 0 & 0 & 0 & 0 & 0 & 0 & 0 & 0 & 0 & 0 & 0 & 0 & 2 & 0 & -2 \\
\end{array}
\begin{array}{c}
),\\
),\\
),\\
),\\
\end{array}
\end{align}
which span the plane $P_{27}$. These vectors have been written in the basis $\{b_\psi\}_{\psi \in \mathcal{Q}}$ with ordering as indicated in Eq.~(\ref{eq:defQ}). To aid the eye, we use vertical lines to demarcate the different species of fermions.

\subsection{Step 4: Testing if an abelian extension sits in a semi-simple algebra} \label{sec:testing}

In this Subsection we explain how to test whether a given abelian extension $\epsilon: \sm_S\oplus \a\rightarrow \su(48)$ sits in the semi-simple extension $\g_i$. Associated with $\epsilon$ is a set of $\u$ charges which we denote $\{X^n\}$, with $X^n\in \mathbb{Q}^{18}$. The abelian extension $\epsilon$ sits in $\g_i$, in the sense of (\ref{eq:triangleAbelien}), if the plane spanned by $\{X^n\}$ sits in $P_i$, up to family permutations.

For each plane $P_i$, whose dimension we label $d_i$, we choose a basis $\{R^{i,j}\}$ and a set of $d_i$ `dual' vectors $\{n^{i,j}\}$, such that  
\begin{align}\label{eq:duality_relations}
R^{i,j}\cdot n^{i,j^\prime}=\delta_{j,j^\prime}.
\end{align}
This then allows us to define a projection $\mathcal{P}_i: \Q^{18}\rightarrow P_i$, where
\begin{align}
\mathcal{P}_i(X)=\sum_{j =1}^{d_i}( n^{i,j}\cdot X) R^{i,j}.
\end{align}
A set of charges $X\in \Q^{18}$ lies in $P_i$ if and only if $\mathcal{P}_i(X)=X$.

Family permutations form a group $S_3^{\times 6}$ which acts on $\Q^{18}$. Denote by $S_i$ the normal subgroup of $S_3^{\times 6}$ that is the stabiliser of  ({\em i.e.} preserves point-wise)  the set $\{n^{i,j}\}$. The plane spanned by $\{X^n\}$ lies in $P_i$ \emph{up to family permutations} if and only if there exists a $[\sigma]\in S_3^{\times 6}/S_i$ and a $\tilde \sigma \in  S_3^{\times 6}$ such that for each $X^n$ 
\begin{align} \label{eq:testing}
\mathcal{P}_i( \sigma(X^n))=\tilde \sigma(X^n).
\end{align} 
Note that the LHS is independent of the choice of representative in the class $[\sigma]$.

The existence of $\tilde \sigma$ can be checked via lexicographic ordering. However, one has to scan through all possible $[\sigma]\in S_3^{\times 6}/S_i$. Thus, for computational efficiency, we wish to find a choice of basis $\{n^{i,j}\}$ such that the stabiliser group $S_i$ is as large as possible. We do this by choosing a subset $\mathcal{Q}_i\subseteq \mathcal{Q}$ of cardinality $d_i$ and defining our $\{n^{i,j}\}=\{b_{\tilde \psi}\}_{\tilde \psi\in \mathcal{Q}_i}$ for which there exists an associated basis  $\{R^{i,\tilde \psi}\}_{\tilde \psi\in \mathcal{Q}_i}$. The elements of $\mathcal{Q}_i$ are chosen so that $\mathcal{Q}_i$ contains as few different species as possible.  In practice, we find both $\mathcal{Q}_i$ and $R^{i,\tilde \psi}$ via row-reduction in matrices with permuted species. One way to interpret the list  $\mathcal{Q}_i$ is as the minimum number of charges one needs to specify a point in $P_i$.

\subsubsection{Example (contd. IV): $\g_{27}=\su(5)^{\oplus 2} \oplus \su(2)^{\oplus 2}$}
We choose $\mathcal{Q}_{27}=\{Q_1,Q_2,Q_3, N_1\}$. This gives $S_{27}\cong (S_3)^{\times 4}\times S_2$, and thus the group $S_3^{\times 6}/S_{27}$ is order $(3!)^6/((3!)^4\times 2!)=18$. The `dual' basis is then given by 
\begin{align}
n^{27,1}=b_{q_1},\quad n^{27,2}=b_{q_2}, \quad n^{27,3}=b_{q_3}, \quad n^{27,4}=b_{\nu^c_1}.
\end{align}
The corresponding $\{R^{i,j}\}$ are 
\begin{align}
\begin{array}{c}
R^{27,1}=\frac{1}{2}(\\
R^{27,2}=\frac{1}{2}(\\
R^{27,3}=\frac{1}{2}(\\
R^{27,4}=\frac{1}{2}(\\
\end{array}
\begin{array}{ccc|ccc|ccc|ccc|ccc|ccc}
 2 & 0 & 0 & -8 & 0 & 0 & 0 & 0 & 4 & 0 & 0 & -6 & 12 & 0 & 0 & 0 & 0 & 0 \\
 0 & 2 & 0 & 0 & -5 & -3 & 2 & 2 & 0 & -3 & -3 & 0 & 0 & 5 & 7 & 0 & 0 & 0 \\
 0 & 0 & 2 & 0 & -3 & -5 & 2 & 2 & 0 & -3 & -3 & 0 & 0 & 7 & 5 & 0 & 0 & 0 \\
 0 & 0 & 0 & 0 & 0 & 0 & 0 & 0 & 0 & 0 & 0 & 0 & 0 & 0 & 0 & 2 & 0 & -2 \\
\end{array}
\begin{array}{c}
)\\
)\\
)\\
)\\
\end{array}
\end{align}
From these it is easy to check that the duality relations in Eq.~(\ref{eq:duality_relations}) hold.

Suppose we want to show that the plane spanned by the vectors $\{Y,Y_3, T_{12}, N_{12}\}$ sits in $P_{27}$, where (see also Table~\ref{tab:charges}) 
\begin{align}\label{eq:Plan27}
\begin{array}{r}
Y:=(\\
Y_3:=(\\
T_{12}:=(\\
N_{12}:=(\\
\end{array}
\begin{array}{ccc|ccc|ccc|ccc|ccc|ccc}
 1 & 1 & 1 & -4 & -4 & -4 & 2 & 2 & 2 & -3 & -3 & -3 & 6 & 6 & 6 & 0 & 0 & 0 \\
 1 & 1 & 0 & -4 & -4 & 0 & 2 & 2 & 0 & -3 & -3 & 0 & 6 & 6 & 0 & 0 & 0 & 0 \\
 0 & 0 & 0 & 0 & 0 & 0 & 0 & 0 & 0 & 0 & 0 & 0 & 0 & 0 & 0 & 1 & -1 & 0 \\
 1 & -1 & 0 & 1 & -1 & 0 & 0 & 0 & 0 & 0 & 0 & 0 & 1 & -1 & 0 & 0 & 0 & 0 \\
\end{array}
\begin{array}{c}
)\\
)\\
)\\
)\\
\end{array}
\end{align}
A scan through all possibilities can be used to find appropriate $[\sigma]\in S_3^{\times 6}/S_{27}$. A suitable choice of permutation class $[\sigma]$ is one which acts on $b_{q_j}$ by sending $(b_{q_1}, b_{q_2}, b_{q_3}) \to (b_{q_3}, b_{q_2}, b_{q_1})$, and sends $b_{\nu^c_2}\rightarrow b_{\nu^c_1}$. As a representative of this class, let $\sigma$ act trivially on all other species (except $b_{q_i}$ and $b_{\nu^c_i}$), and send $(b_{\nu^c_1}, b_{\nu^c_2}, b_{\nu^c_3}) \to (b_{\nu^c_2}, b_{\nu^c_1}, b_{\nu^c_3})$.  Namely, we choose the representative 
\begin{align}
\begin{array}{r}
\sigma(Y)=(\\
\sigma(Y_3)=(\\
\sigma(T_{12})=(\\
\sigma(N_{12})=(\\
\end{array}
\begin{array}{ccc|ccc|ccc|ccc|ccc|ccc}
 1 & 1 & 1 & -4 & -4 & -4 & 2 & 2 & 2 & -3 & -3 & -3 & 6 & 6 & 6 & 0 & 0 & 0 \\
 0 & 1 & 1 & -4 & -4 & 0 & 2 & 2 & 0 & -3 & -3 & 0 & 6 & 6 & 0 & 0 & 0 & 0 \\
 0 & 0 & 0 & 0 & 0 & 0 & 0 & 0 & 0 & 0 & 0 & 0 & 0 & 0 & 0 & -1 & 1 & 0 \\
 0 & -1 & 1 & 1 & -1 & 0 & 0 & 0 & 0 & 0 & 0 & 0 & 1 & -1 & 0 & 0 & 0 & 0 \\
\end{array}
\begin{array}{c}
)\\
)\\
)\\
)\\
\end{array}
\end{align}
We can then act on these with the projection matrices
\begin{align}\label{eq:Proj27}
\begin{array}{r}
\mathcal{P}_{27}(\sigma(Y))=(\\
\mathcal{P}_{27}(\sigma(Y_3))=(\\
\mathcal{P}_{27}(\sigma(T_{12}))=(\\
\mathcal{P}_{27}(\sigma(N_{12}))=(\\
\end{array}
\begin{array}{ccc|ccc|ccc|ccc|ccc|ccc}
 1 & 1 & 1 & -4 & -4 & -4 & 2 & 2 & 2 & -3 & -3 & -3 & 6 & 6 & 6 & 0 & 0 & 0 \\
 0 & 1 & 1 & 0 & -4 & -4 & 2 & 2 & 0 & -3 & -3 & 0 & 0 & 6 & 6 & 0 & 0 & 0 \\
 0 & 0 & 0 & 0 & 0 & 0 & 0 & 0 & 0 & 0 & 0 & 0 & 0 & 0 & 0 & -1 & 0 & 1 \\
 0 & -1 & 1 & 0 & 1 & -1 & 0 & 0 & 0 & 0 & 0 & 0 & 0 & 1 & -1 & 0 & 0 & 0 \\
\end{array}
\begin{array}{c}
)\\
)\\
)\\
)\\
\end{array}
\end{align}
Comparing Eq.~(\ref{eq:Plan27}) with Eq.~(\ref{eq:Proj27}), it is easy to see that a $\tilde \sigma \in (S^3)^{\times 6}$ exists for which 
\begin{align}
\mathcal{P}_{27}(\sigma(Y))=\tilde \sigma(Y), \quad \mathcal{P}_{27}(\sigma(Y_3))=\tilde \sigma(Y_3),\quad \mathcal{P}_{27}(\sigma(T_{12}))=\tilde \sigma(T_{12}),\quad \mathcal{P}_{27}(\sigma(N_{12}))=\tilde \sigma(N_{12}).
\end{align}
Thus, up to equivalence under family permutations, the plane $\mathrm{Span}(Y,Y_3, T_{12},N_{12})$ lies within $P_{27}$. Moreover, since the dimension of both planes is $4$ they must coincide. Hence, we may write
\begin{align}\label{eq:P27T}
P_{27} = \mathrm{Span}(Y,Y_3, T_{12},N_{12})\, .
\end{align}
As we soon see, {\em all} the planes in Flatland can be similarly represented as a span of some of the vectors appearing in Table~\ref{tab:charges}.

\subsubsection{Examples of permutation invariant diagnostics}
For certain planes in our list a short predicate (a true-or-false statement) can be contrived specifying if a given $X$ sits in a plane up to permutation.
Two pertinent examples where this is the case are the planes $P_4$ and $P_6$, which recall correspond to the gauge algebras 
$\g_4 = \su(4)\oplus \sp(6)^{\oplus 2}$ and $\g_6=\su(12) \oplus \su(2)^{\oplus 2}$,
the predicates for which we study here. The planes $P_4$ and $P_6$ can be written in a similar fashion to Eq.~(\ref{eq:P27T}), as
\begin{align} \label{eq:P4}
P_4&=\mathrm{Span}(Y,BL,S_{12}, D_{12}, D_{23}),\\
P_{6}&=\mathrm{Span}(Y,BL,B_{12},B_{23},L_{12},L_{23})\, ,
\label{eq:P6T}
\end{align}
where the sets of charges appearing in this equation are given in Table~\ref{tab:charges}.

We require a little more notation to continue, as follows. Given a set of charges $X$ satisfying the ACCs, we let $X_F$ for each $F\in \{q,u^c,d^c,\ell,e^c,\nu^c\}$ denote a diagonal $3\times 3$ matrix with diagonal entries equal to $X_{F_j}$, $j\in\{1,2,3\}$. The values of $\Tr X_F$, $\Tr X_F^2$, and $\det X_F$ determine these charges uniquely up to permutation. We can moreover trade each $X_F$ for a traceless matrix $\tilde X_F$, by the addition of some amount of $BL$ and $Y$. Specifically, we define 
\begin{align} \label{eq:Xtilde}
\tilde X_F=3X_F+\frac{1}{3} ( \Tr X_q+ \Tr X_{u^c}) Y_F - \frac{1}{3} (4 \Tr X_q+\Tr X_{u^c}) BL_F.
\end{align}
where by construction $\Tr\tilde X_F=0$.

From Eq.~(\ref{eq:Xtilde}) and Eq.~(\ref{eq:P4}), it is clear that $X\in P_4$ (up to permutation) if and only if $\tilde X$ is in $\mathrm{Span}(S_{12},D_{12},D_{23})$ up to permutation. Let us focus on the species $L$ and $Q$, we must have that 
\begin{align}
\tilde X_{\ell}=\tilde X_{q}=a (S_{12})_q
\end{align}
for some $a\in \mathbb{Z}$, and up to permutation. This holds if and only if 
\begin{align}
\det \tilde X_\ell=\det\tilde X_q=a^3 \det (S_{12})_q=0\, ,\\
\Tr \tilde X_\ell^2=\Tr \tilde X_q^2=a^2 \Tr (S_{12})_q^2=2 a^2\, ,
\end{align}
for some $a\in \mathbb{Z}$. Since $\Tr\tilde X_\ell=0$ and $\Tr\tilde X_q=0$, these conditions are equivalent to 
\begin{align}
\det \tilde X_\ell&=\det\tilde X_q=0\, ,\\
\Tr \tilde X_\ell^2&=\Tr \tilde X_q^2\, .
\end{align}
A similar argument can be applied for the fields $d^c,e^c,\nu^c$ and $u^c$, which gives us the predicate
\begin{align}
\mathbb{P}_4:=(\quad & \det \tilde X_\ell=\det \tilde X_q=0\nonumber \\ 
\wedge \quad & \Tr \tilde X_q^2=\Tr \tilde X_\ell^2\nonumber \\
\wedge \quad & \det \tilde X_{d^c}=\det \tilde X_{e^c}=-\det \tilde X_{\nu^c} =-\det \tilde X_{u^c}\nonumber \\
\wedge \quad & \Tr \tilde X_{d^c}^2=\Tr \tilde X_{e^c}^2=\Tr \tilde X_{\nu^c}^2=\Tr \tilde X_{u^c}^2)
\end{align}
Likewise, for the plane $P_6$ we get
\begin{align}
\mathbb{P}_6:=(\quad & \det \tilde X_{e^c}=\det \tilde X_{\nu^c}=-\det \tilde X_\ell\nonumber \\
\wedge \quad & \Tr \tilde X_{e^c}^2=\Tr \tilde X_{\nu^c}^2=\Tr \tilde X_{\ell}^2\nonumber \\
\wedge \quad & \det \tilde X_{u^c}=\det \tilde X_{d^c}=-\det \tilde X_q\nonumber \\
\wedge \quad & \Tr \tilde X_{u^c}^2=\Tr \tilde X_{d^c}^2=\Tr \tilde X_q^2)
\end{align}
These predicates could be used to test whether a set of charges sits in $P_4$ or $P_6$ very efficiently, without having to cycle through permutations in the manner described in Subsection~\ref{sec:testing}.

\section{Results} \label{sec:results}

Having described our method in detail in the previous two Sections, we here summarise our results. 
We obtain the planes of $X$-charge assignments which {parametrise all abelian extensions of the SM} that embed in each of the $27$ maximal algebras listed in Table~\ref{tab:maximal}.
The planes are, however, not independent from one another. 

We thus identify planes that are equal, as follows:
\begin{align}
P_4\sim P_5\sim P_{19}\sim P_{20} \sim P_{21} \quad [ P_4]\, ,\nonumber \\
P_8\sim P_9\sim P_{10}\sim P_{11}\sim P_{12}\sim P_{13}\quad [ P_8]\, ,\nonumber \\
P_{7}\sim P_{17}\sim P_{18}\sim P_{24}\quad [ P_7]\, ,
\end{align}
where the plane in parentheses at the end of each line indicates how we denote the whole equivalence class. All other planes sit in their own equivalence class. 

\begin{figure}
\begin{center}
\includegraphics[width=3in]{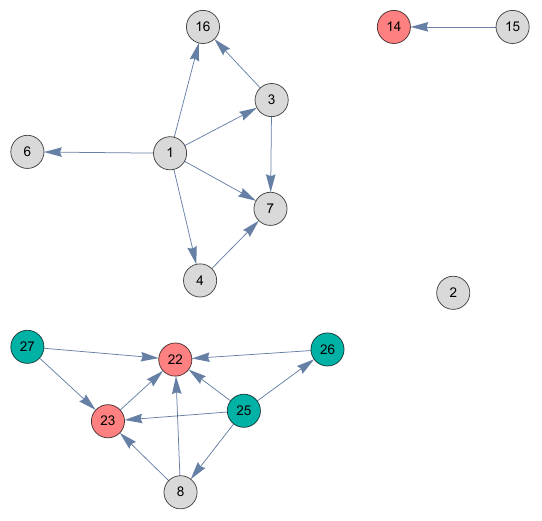}
\caption{{The interrelationships between the different (equivalence classes of) planes. An arrow from $i$ to $j$ indicates that $P_i$ is contained entirely within $P_j$ (up to family permutations). If the dot associated with $i$ is coloured {\color{myred}red} then every $\g_k$ with plane $P_k$ equivalent to $P_i$ has global anomalies. If the dot is coloured {\color{mygreen} green} then every $\g_k$ with plane $P_k$ equivalent to $P_i$ is only maximal once those $\g_i$'s with global anomalies are removed (\emph{i.e.} $k=25$, $26$ or $27$).} \label{fig:graph}}
\end{center}
\end{figure}

Moreover, some (equivalence classes of) planes sit entirely within others.
These relationships between planes are summarised in Fig.~\ref{fig:graph}, and can be used to deduce a minimal set of planes that need to be considered ({\em i.e.} such that everything embeds in at least one of these planes, but no two of these planes embed in one another). To give one example, the plane $P_4$, that parametrises abelian extensions that embed in the electroweak flavour unification algebra $\g_4=\su(4)\oplus \sp(6)_L \oplus \sp(6)_R$, itself embeds in the plane $P_7$ associated to $\g_7=\su(4)\oplus \sp(4)_L \oplus \sp(4)_R \oplus \so(10)$. This relationship is represented in Fig.~\ref{fig:graph} by the (directed) arrow connecting nodes labelled `$4$' and `$7$'. 

The following planes (\ref{eq:P2}--\ref{eq:P16}) need to be considered,
whether or not we are accounting for global gauge anomalies:
\begin{align}
P_{2}&=\mathrm{Span}(Y_1,Y_2,Y_3,BL_1,BL_2,BL_3)\, , \label{eq:P2}\\
P_{6}&=\mathrm{Span}(Y,BL,B_{12},B_{23},L_{12},L_{23})\, , \label{eq:P6}\\
P_{7}&=\mathrm{Span}(Y,BL,Y_{3},BL_{3}, S_{12},D_{12})\, ,\\
P_{16}&=\mathrm{Span}(Y,BL,Y_3,BL_{3},L_{12}, B_{12})\, . \label{eq:P16}
\end{align} 
Recall from Section \ref{sec:global} that the list of maximal gauge algebras changes depending on whether we account for global anomaly cancellation or not. If we ignore global anomalies, the following planes (\ref{eq:P14}--\ref{eq:P22}) should be considered, in addition to  (\ref{eq:P2}--\ref{eq:P16}),
\begin{align}
P_{14}&=\mathrm{Span}(Y,Y_{3},BL_{3}, Y_1, N_{12})\, , \label{eq:P14}\\
P_{22}&=\mathrm{Span}(Y,Y_{3},BL_{3},F_{12},T_{12},N_{12})\, . \label{eq:P22}
\end{align}
If, however, we restrict to gauge algebras for which there are possible gauge groups that are free of local {\em and} global anomalies, then the algebras $\g_{14}$ and $\g_{22}$ (amongst others) are struck out, and other $\g_i\in\S$ become maximal as described in Section \ref{sec:global}. 
In this case, accounting for global anomalies, the following planes (\ref{eq:P8}--\ref{eq:P27}) need also be considered, in place of (\ref{eq:P14}--\ref{eq:P22}):
\begin{align}
P_{8}&=\mathrm{Span}(Y, F_{12}, T_{12},N_{12}) \, ,\label{eq:P8} \\
P_{15}&=\mathrm{Span}(Y_1,Y_2,Y_3, N_{12})\, ,\\
P_{26}&=\mathrm{Span}(Y,Y_{3},BL_3, F_{12}+N_{12},T_{12})\, ,\\
P_{27}&=\mathrm{Span}(Y,Y_{3},T_{12},N_{12})\, . \label{eq:P27}
\end{align}
If a given anomaly-free {set of $\u_{X^1} \oplus \u_{X^2}\oplus \dots$ charge assignments}
sits in one of these planes, then this abelian extension of the SM embeds inside the corresponding $\g_i$ in that equivalence class. If a $X$-charge assignment sits in none of these planes, then it has no semi-simple completion. 

We emphasise that these planes $P_i$ parametrise the abelian SM extensions $\epsilon:\sm_S \oplus \a \to \su(48)$ which sit in semi-simple extensions of the SM, as introduced carefully in Section~\ref{sec:formalism}.{ We refer to the collection of planes, that covers all cases of abelian extensions with semi-simple completions, as `Flatland'. As explained, we distinguish two cases, in which we consider completions which are free of only local anomalies ($\mathcal{F}^\prime$), and completions which are free of both local and global anomalies ($\mathcal{F}$), which are:
\begin{align}
\mathcal{F}^\prime&:=(P_2 \cup P_6\cup P_7 \cup P_{16})\cup (P_{14}\cup P_{22})\label{eq:FLLA}\,,\\
\mathcal{F}&:=(P_2 \cup P_6\cup P_7 \cup P_{16})\cup (P_{8}\cup P_{15}\cup P_{26}\cup P_{27})\label{eq:FLGALA}\,.
\end{align}

As explained {in Section~\ref{sec:formalism}}, the abelian part $\a$ necessarily includes the generator of hypercharge that will remain unbroken (as can indeed be seen from the formulae (\ref{eq:P2}--\ref{eq:P27}) for the 8 planes), thanks to our requiring the embedding $\iota$ in (\ref{eq:iota}). If we interpret such a gauge model $\sm_S \oplus \a$ as an extension of the SM, the idea is that $\sm_S \oplus \a$ breaks to $\sm$. For each plane $P_i$, the maximum number of independent $\ZP$ gauge bosons that will be produced is thus
\begin{equation}
\mathrm{Dim}(\sm_S \oplus \a) - \mathrm{Dim}(\sm) \leq d_i - 1\, .
\end{equation}
Since the dimension of the planes is $\leq 6$, we conclude that the largest number of independent $\ZP$ bosons we can produce from a semi-simple extension is 5.}

\subsection{Vector-like solutions and the plane $P_6$} \label{sec:VL}
An important subclass of anomaly-free $\sm \oplus \u_{X^1}\oplus \u_{X^2}\oplus \dots$ theories is given by those where each $X^n$ charge assignment is {\em vector-like}, which means that left- and right-handed fields are charged equally.\footnote{More precisely, a set of charges $X$ is vector like if and only if the sets of charges $\{Q_1,Q_2,Q_3\}$, $\{-U_1,-U_2,-U_3\}$, and $\{-D_1,-D_2, -D_3\}$ contain identical elements, and likewise for the sets of charges $\{L_1,L_2,L_3\}$, $\{-E_1,-E_2,-E_3\}$, and $\{-N_1,-N_2, -N_3\}$.} From (\ref{eq:P6}), it is easy to see that every set of charges in $X\in P_6$ is vector-like, up-to the addition of hypercharge, where recall that the gauge algebra $\g_6 = \su(12)\oplus \su(2)\oplus \su(2)$. 
As we show shortly, an inverted statement is also true, namely\footnote{The fact that a class of such anomaly-free vector-like $\sm \oplus \u_X$ models embeds in $\su(12) \oplus \su(2) \oplus \su(2)$ was appreciated in Ref.~\cite{Davighi:2022qgb}.} 
\be \label{eq:InvertedStatement}
\text{\emph{Every anomaly-free vector-like extension $\sm \oplus \u_{X^1}\oplus \u_{X^2}\oplus \dots$ sits in  $\g_6$.}}
\ee 

These results are not that surprising; the algebra $\g_6$ fully unifies 3-family flavour symmetry with the quark-lepton unified $\su(4)$ colour symmetry of Pati--Salam~\cite{Pati:1974yy}, and colour is vector-like. This should be contrasted, for example, with the corresponding plane $P_4$ for the electroweak flavour unification algebra $\g_4$; in that case, where flavour is unified with {\em chiral} electroweak symmetries, we get a plane $P_4$ containing many chiral `directions', namely $S_{12}$, $D_{12}$, and $D_{23}$. 

To see that our inverted statement~(\ref{eq:InvertedStatement}) is true,  first note that a general vector-like $X$ can be parametrised as a linear combination of family-specific baryon and lepton numbers 
\begin{equation} \label{eq:VLX}
X = \sum_{i=1}^3 \left(\beta_i B_i + \lambda_i L_i \right)\, ,
\end{equation}
for coefficients $\{\beta_i, \lambda_i\} \in \Q^6$. The fact that $X$ is vector-like means that the cubic anomaly, the gravitational anomaly, and the mixed anomaly with $\su(3)$ all vanish automatically. Substituting (\ref{eq:VLX}) into the $\sm\oplus \u_X$ anomaly cancellation equations, one finds that the mixed anomaly involving one hypercharge boson and two $\u_X$ bosons vanishes also, even though hypercharge is chiral. This leaves only a pair of mixed anomalies between $\u_X$ and $\su(2)\oplus \u_Y$ that are linear in the $X$ charges, and both these anomaly coefficients are proportional to 
$\sum_{i=1}^3 (\beta_i + \lambda_i)$, which must therefore vanish. Thus, the most general anomaly-free vector-like $X$ corresponds to gauging a linear combination of baryon and lepton numbers, with coefficients that sum to zero. This tells us that a generic such $X$ must sit in the plane $P_6$.

Turning to a generic anomaly-free vector-like extension $\sm \oplus \u_{X^1}\oplus \u_{X^2}\oplus \dots$, our above results tells us that, since each $X^n$ is vector-like and $\sm\oplus \u_{X^n}$ is anomaly free, each $X^n$ must lie in $P_6$ (we note that $P_6$ is invariant under family permutations). Thus, the span $\{Y,X^n\}$ forms a sub-plane of $P_6$ and is automatically anomaly free. This leads us directly to the result in~(\ref{eq:InvertedStatement}). But also the further statement, that any extension of the $\sm$ by any number of vector-like charges is guaranteed to be anomaly free, provided each $\sm\oplus \u_{X^n}$ is anomaly free (without having to check any of the extra mixed anomalies explicitly).

\subsection{Testing a selection of known models} \label{sec:examples}

Given any particular anomaly-free abelian extension of the SM, of which examples abound in the literature (especially featuring a single $\u_X$ extension), using our method we can easily ascertain whether the model embeds in a semi-simple gauge algebra $\g \in \S$. We can moreover identify which equivalence classes of planes it embeds in, from the list (\ref{eq:P2}--\ref{eq:P27}). We collect results for a selection of models from the literature in Table~\ref{tab:models2}. {The tests of all these models are included in the computer program \texttt{Test\_you\_own\_charges.nb} that is included with this \texttt{arXiv} submission, and that we describe how to use in Section~\ref{sec:program}.}

We introduce the notation $U_{i_1, i_2, \dots i_n}$ to indicate the set of solutions in the planes $P_{i_1}$, $P_{i_2}$, $\dots$ $P_{i_n}$ but not in any others in the list  (\ref{eq:P2}--\ref{eq:P27}). In particular, $U_{\emptyset}$ is the set of solutions that do not embed in any of the planes, corresponding to $\sm \oplus \u_X$ extensions that do not have any semi-simple gauge completions.

To draw out some notable results, we see that all vector-like extensions, including the rank-3 $\u_{L_e-L_\mu} \oplus \u_{L_e-L_\tau} \oplus \u_{L_\mu-L_\tau}$ extension, embed at least in $P_6$ in accordance with Subsection~\ref{sec:VL}. This includes various models that have been used to explain the $b\to s\mu\mu$ anomalies, such as models based on $B_3-L_2$~\cite{Allanach:2020kss,Alonso:2017uky,Bonilla:2017lsq} (which embeds in many more planes besides $P_6$), on  $B-3L_2$~\cite{Greljo:2021xmg}, and its class of generalisations from Ref.~\cite{Davighi:2022qgb} that also feature exact proton stability. The Third Family Hypercharge ($Y_3$) Model~\cite{Allanach:2018lvl} and the $X=Y_3-3(B-L)_3$ variant~\cite{Allanach:2022bik}, despite being chiral, embed in many different planes. Other chiral models, such as the $DY_3$ model of~\cite{Allanach:2019iiy}, the chiral muoquark models of~\cite{Greljo:2021npi,Greljo:2022dwn}, and the $\sm\oplus \u^{\oplus 2}$ `anti-GUT inspired' neutrino mass model of~\cite{Froggatt:1998he},\footnote{The charges of the anomaly-free $\u^{\oplus 2}$ extension of the SM in~\cite{Froggatt:1998he} were motivated by an underlying $\sm^{\oplus 3} \oplus \u$ `anti grand unification' model. 
The fact that this model embeds in the anti-GUT, which is not semi-simple, does not of course contradict our finding that this model has no semi-simple completions.
} do not sit in any of our planes, and so none of these chiral models have a semi-simple completion.

\begin{table}
\begin{adjustwidth}{-.7in}{-.7in}  
\begin{equation*}
\begin{array}{|c|c|ccc|ccc|ccc|ccc|ccc|ccc|c|}
\hline
\text{Model} & \text{Refs} & q_1 & q_2 & q_3 & u_1^c& u_2^c & u_3^c & d_1^c & d_2^c & d_3^c & \ell_1 & \ell_2 & \ell_3 &e_1^c & e_2^c &e_3^c & \nu_1^c &\nu_2^c &\nu_3^c & \g \text{~planes} \\ \hline
B-L & \text{\cite{PhysRevLett.44.1316,PhysRevD.20.776}} & 1&1&1&-1&-1&-1&-1&-1&-1&-3&-3&-3&3&3&3&3&3&3& U_{2,6,7,16}\\
Y_3 & \text{\cite{Allanach:2018lvl}} & 0&0&1& 0&0&-4& 0&0&2&0&0&-3&0&0&6&0&0&0&U_{2,7,16,14,22,15,26,27} \\
DY_3 & \text{\cite{Allanach:2019iiy}} & 0&0&1& 0&0&-4& 0&0&2&0&5&-8&0&-4&10&0&0&0&U_{\emptyset} \\
L_\mu-L_\tau \text{~(I)} & \text{\cite{He:1991qd}} & 0&0&0&0&0&0&0&0&0&0&1&-1&0&-1&1&0&0&0&U_{\emptyset}\\
L_\mu-L_\tau \text{~(II)} & \text{\cite{He:1991qd}} & 0&0&0&0&0&0&0&0&0&0&1&-1&0&-1&1&0&-1&1& U_{6,16}\\
\tilde{L}_{\mu-\tau} & \text{\cite{Greljo:2021npi,Greljo:2022dwn}} & 0&0&0&0&0&0&0&0&0&-1&7&6&-1&-6&7&7&2&-9&U_{\emptyset}\\
B_3-L_2 & \text{\cite{Allanach:2020kss}} &0&0&1&0&0&-1&0&0&-1&0&-3&0&0&3&0&0&3&0& U_{2,6,7,16,14,22,26}\\
Y_3-3(B-L)_3 & \text{\cite{Allanach:2022bik}} & 0&0&-2& 0&0&-1& 0&0&5&0&0&6&0&0&-3&0&0&-9&U_{2,7,16,14,22,26} \\
B-3L_2 & \text{\cite{Greljo:2021xmg}} &1&1&1&-1&-1&-1&-1&-1&-1&0&-9&0&0&9&0&0&9&0& U_{6}\\
 \hline
 \{BL,Y_3\} & & 1 & 1 & 1 & -1 & -1 & -1 & -1 & -1 & -1 & -3 & -3 & -3 & 3 & 3 & 3 & 3 & 3 & 3 & U_{2,7,16}\\
 & & 0 & 0 & 1 & 0 & 0 & -4 & 0 & 0 & 2 & 0 & 0 & -3 & 0 & 0 & 6 & 0 & 0 & 0 & \\ 
 \hline 
 \text{`anti-GUT'} & \text{\cite{Froggatt:1998he}} & 0 & 0 & 1 & 0 & -4 & 0 & 0 & 0 & 2 & 0 & 0 & -3 & 0 & 0 & 6 & 0 & 0 & 0 & U_{\emptyset}\\
  & & 0 & -1 & 0 & 0 & 1 & 3 & 0 & -1 & -1 & 0 & 3 & 0 & 0 & -5 & -1 & 0 & 0 & 0 & \\ 
 \hline 
 \{L_e-L_\mu, & & 0 & 0 & 0 & 0 & 0 & 0 & 0 & 0 & 0 & 1 & -1 & 0 & -1 & 1 & 0 & -1 & 1 & 0 & U_{6}\\
  L_e-L_\tau, & & 0 & 0 & 0 & 0 & 0 & 0 & 0 & 0 & 0 & 1 & 0 & -1 & -1 & 0 & 1 & -1 & 0 & 1 & \\
  L_\mu-L_\tau\} & &  0 & 0 & 0 & 0 & 0 & 0 & 0 & 0 & 0 & 0 & -1 & 1 & 0 & 1 & -1 & 0 & 1 & -1 & \\
  \hline
\end{array}
\end{equation*}
\end{adjustwidth}
\caption{Selection of anomaly-free $U(1)_X$ extensions of the SM, and the planes in which they embed. The notation in the far column $U_{i_1, \dots i_n}$ indicates that the given charge assignment embeds in the planes $P_{i_1}$, $P_{i_2}$, $\dots$ $P_{i_n}$ but not in any others in the list  (\ref{eq:P2}--\ref{eq:P27}). Embedding in a plane $P_{i}$ means that the $\sm \oplus \u_X$ extension embeds in a semi-simple gauge theory with gauge algebra $\g_i$, corresponding to the numbered list in Table~\ref{tab:maximal} (as well as other algebras for which the planes are equivalent, as explained in the main text). 
In addition to the particular solution labelled $\tilde{L}_{\mu-\tau}$, we tested all the other 20 `chiral muoquark' models from~\cite{Greljo:2021npi,Greljo:2022dwn}, finding that all are in $U_{\emptyset}$, meaning that none of these $\sm \oplus \u_X$ models have semi-simple completions. In the bottom three rows, we also include examples of multi-$\ZP$ extensions, specifically two $\u^{\oplus 2}$ extensions and one $\u^{\oplus 3}$ extension, as a proof of principle.
\label{tab:models2} }    
\end{table}

\subsection{Comparison with the anomaly-free atlas}\label{sec:Atlas}
In this section, we investigate `what fraction' of anomaly-free $\u_X$ extensions of the SM$+3\nu_R$ can be embedded in {\em any} anomaly-free semi-simple extension of the SM$+3\nu_R$. For a set of charges $X\in \mathbb{Q}^{18}$ {(that we label $Q_j,U_j,D_j,L_j,E_j$ and $N_j$)} to be anomaly-free it must satisfy the ACCs, which are a set of Diophantine equations:
\begingroup
\allowdisplaybreaks
\begin{align}
0&=\sum_{j=1}^3(6Q_j+3U_j+3D_j+2L_j+E_j+N_j)\,,\\
0&=\sum_{j=1}^3(3Q_j+2L_j)\,,\\
0&=\sum_{j=1}^3(2Q_j+U_j+D_j)\,,\\
0&=\sum_{j=1}^3(Q_j+8U_j+2D_j+3L_j+6E_j)\,,\\
0&=\sum_{j=1}^3(Q_j^2-2U_j^2+D_j^2-L_j^2+E_j^2)\,,\\
0&=\sum_{j=1}^3(6Q_j^3+3U_j^3+3D_j^3+2L_j^3+E_j^3+N_j^3)\,.
\end{align}
\endgroup
Due to the \emph{non}-linear quadratic and cubic equations, the total space of anomaly-free $X$ charges in $\mathbb{Q}^{18}$ is complicated. An analytic parametrisation of this space is given in Ref.~\cite{Allanach:2020zna}. This followed a numerical scan in Ref.~\cite{Allanach:2018vjg} of points up to some maximum height of $\Qmax=10$. Here $\Qmax$ is defined as the maximum absolute value of a set of charges $X$, once it has been rescaled to contain co-prime integers. (Note that two sets of charges that differ by family permutations or by a constant rescaling should be considered `the same'.)

Our planes define linear subspaces of the overall space of anomaly-free $X$ charges. We can perform the task of determining how many solutions up to a given $\Qmax$ lie in each plane. There are two ways to do this. Firstly we can scan through the lists of solutions provided in Ref.~\cite{Allanach:2018vjg} and test, which lie in our planes. Secondly, we can generate these solutions explicitly by scanning through permissible coefficients of the vectors $\{R^{i,j}\}$. As a check, we perform both analyses. 

By using the $U_{i_1,i_2,\ldots,i_n}$ notation introduced in the previous Subsection, we can give complete information, short of providing the solutions themselves, about how many (inequivalent) solutions sit in each of the 10 planes up to $\Qmax=10$. The results are:
\begin{align}
\begin{small}
\begin{array}{llllll}
|U_{\emptyset}|=20967127 ,  &  |U_{2 , {\color{myred} 14}}|=49 ,  &  |U_{{\color{myred} 22} , {\color{mygreen} 26}}|=6284 ,  &  |U_{2 , {\color{myred} 14} , 16}|=11 ,  &  |U_{2 , 7 , 16 , {\color{myred} 22}}|=8 , \\|U_{2}|=14429 ,  &  |U_{2 , 16}|=3144 ,  &  |U_{{\color{myred} 22} , {\color{mygreen} 27}}|=180 ,  &  |U_{2 , 16 , {\color{myred} 22}}|=19 ,  &  |U_{7 , 16 , {\color{myred} 22} , {\color{mygreen} 26}}|=563 , \\|U_{6}|=52449 ,  &  |U_{2 , {\color{myred} 22}}|=126 ,  &  |U_{2 , 6 , 16}|=83 ,  &  |U_{6 , 7 , 16}|=1147 ,  &  |U_{{\color{mygreen} 8} , {\color{myred} 22} , {\color{mygreen} 26} , {\color{mygreen} 27}}|=5 , \\|U_{7}|=270896 ,  &  |U_{6 , 16}|=9219 ,  &  |U_{2 , 7 , {\color{myred} 14}}|=45 ,  &  |U_{{\color{mygreen} 8} , {\color{myred} 22} , {\color{mygreen} 26}}|=94 ,  &  |U_{{\color{myred} 14} , {\color{mygreen} 15} , {\color{myred} 22} , {\color{mygreen} 27}}|=30 , \\|U_{{\color{myred} 14}}|=1080 ,  &  |U_{7 , 16}|=15563 ,  &  |U_{2 , 7 , 16}|=3869 ,  &  |U_{{\color{mygreen} 8} , {\color{myred} 22} , {\color{mygreen} 27}}|=103 ,  &  |U_{6 , 7 , 16 , {\color{myred} 22} , {\color{mygreen} 26}}|=38 , \\|U_{16}|=113316 ,  &  |U_{{\color{mygreen} 8} , {\color{myred} 22}}|=1251 ,  &  |U_{2 , 7 , {\color{myred} 22}}|=59 ,  &  |U_{{\color{myred} 22} , {\color{mygreen} 26} , {\color{mygreen} 27}}|=18 ,  &  |U_{{\color{mygreen} 8} , {\color{myred} 14} , {\color{mygreen} 15} , {\color{myred} 22} , {\color{mygreen} 27}}|=11 , \\|U_{{\color{myred} 22}}|=70909 ,  &  |U_{{\color{myred} 14} , {\color{mygreen} 15}}|=10 ,  &  |U_{2 , {\color{mygreen} 8} , {\color{myred} 22}}|=3 ,  &  |U_{2 , 6 , 7 , 16}|=139 ,  &  |U_{2 , 7 , {\color{myred} 14} , 16 , {\color{myred} 22} , {\color{mygreen} 26}}|=88 , \\|U_{2 , 7}|=13521 ,  &  |U_{{\color{myred} 14} , {\color{myred} 22}}|=1015 ,  &  |U_{2 , {\color{myred} 14} , {\color{mygreen} 15}}|=1 ,  &  |U_{2 , 7 , {\color{myred} 14} , 16}|=3 ,  &  |U_{6 , 7 , {\color{mygreen} 8} , 16 , {\color{myred} 22} , {\color{mygreen} 26}}|=5 , \\
 \multicolumn{2}{c}{|U_{2 , 7 , {\color{myred} 14} , {\color{mygreen} 15} , 16 , {\color{myred} 22} , {\color{mygreen} 26} , {\color{mygreen} 27}}|=3,}  & \multicolumn{3}{c}{|U_{2 , 6 , 7 , {\color{mygreen} 8} , {\color{myred} 14} , {\color{mygreen} 15} , 16 , {\color{myred} 22} , {\color{mygreen} 26} , {\color{mygreen} 27}}|=2,} \\
\end{array}
\end{small}
\end{align}
~\\
\noindent where we have written in {\color{myred} red} those planes which should only be considered when we are ignoring the effects of global anomalies, and we have written in {\color{mygreen} green} those planes which are only `maximal' when one is taking account of global anomalies. There are exactly two solutions in all of our planes, which unsurprisingly, correspond to hypercharge and the zero-solution.

\begin{figure}
\begin{center}
\includegraphics[width=3.5in]{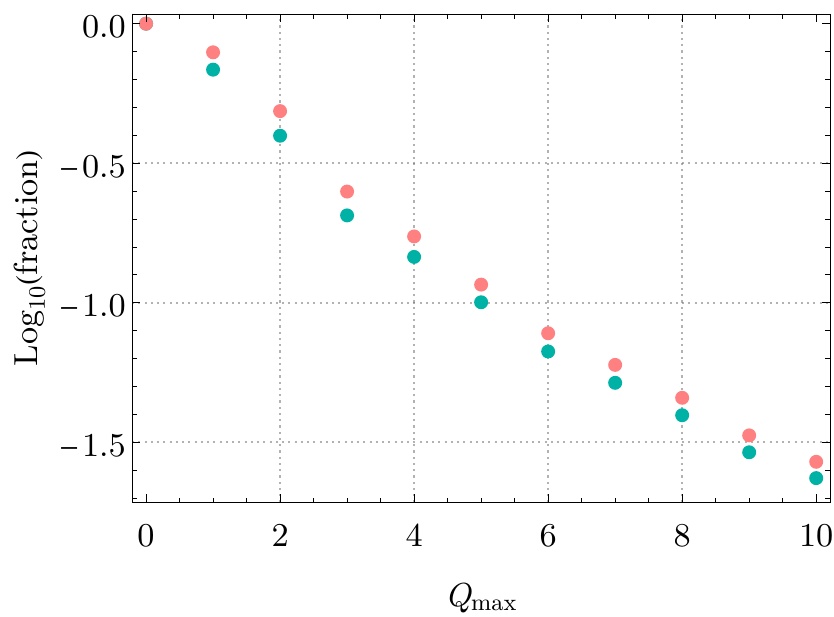}
\caption{The fraction of anomaly-free $\u_X$ extensions of the SM, including three right-handed neutrinos, that embed in any semi-simple gauge model. This is plotted as a function of $\Qmax$, the maximum absolute charge in the solution X after clearing common divisors. The {\color{myred} red} points consider embedding each $\sm\oplus \u_X$ in any $\g$ that is free of local anomalies, while the {\color{mygreen} green} points restrict to $\g$ for which there is a corresponding gauge {\em group} that is further free of {\em global} anomalies. We see that, out to $\Qmax=10$, to which the anomaly-free atlas was charted in~\cite{Allanach:2018vjg}, roughly 2.5 \% of solutions are anomaly-free.} \label{fig:AtlasSummary}
\end{center}
\end{figure}

In Figure~\ref{fig:AtlasSummary} we plot the fraction of all anomaly-free sets of charges that sit in \emph{any} semi-simple extension, 
as a function of $\Qmax$. We summarise the same data in a table below the plot.
The {\color{myred} red} points are for the list of semi-simple algebras \emph{including} those that suffer from global anomalies.
The {\color{mygreen} green} points are for the list of semi-simple algebras when care is taken to \emph{exclude} those which necessarily suffer from global anomalies (see Section \ref{sec:global}).
Since there are more planes the former can sit in, the red points are always higher than the green points. The total fraction of anomaly-free solutions that have semi-simple completions is, for $\Qmax=10$, around $2.5\%$ in both cases. 

\section{Testing your own $\u_{X^1}\oplus \u_{X^2}\oplus \dots$ extension of the SM}  \label{sec:program}
As part of the ancillary directory to the \texttt{arXiv} submission of this paper, we have included several computer programs. These are described below: 

\subsection{Test\_your\_own\_charges.nb} The \texttt{Mathematica} program \texttt{Test\_your\_own\_charges.nb} in our ancillary directory contains a function \texttt{testCharges} that takes as an input a single set of charges $X$, or multiple sets of charges $\{X^n\}$. 

For a general single set of charges $X$, {with all Weyl fermions of the SM+$3\nu_R$ taken to be left-handed, the function is called as follows}
\begin{equation}
\verb!testCharges[{! Q_1,Q_2,Q_3, U_1,U_2,U_3, D_1,D_2,D_3, L_1, L_2, L_3, E_1, E_2, E_3, N_1, N_2,N_3\verb!}]!\, 
\end{equation}
As a specific example, {to test the $Y_3$ assignment as listed in Table~\ref{tab:charges}, one would input}
\begin{equation}\label{eq:ChargeExampleY3}
\verb!testCharges[{!0, 0, 1, 0, 0, -4, 0, 0, 2, 0, 0, -3, 0, 0, 6, 0, 0, 0\verb!}]!\, 
\end{equation}
{noting the sign convention for the charges of the right-handed fields.}

For multiple sets of charges $\{X^n\}$ which span a plane in $\mathbb{Q}^{18}$, the function is called as follows for \emph{e.g.} 2 sets of charges 
\begin{equation*}
\verb!testCharges[{{! Q_1^1,Q_2^1,Q_3^1, U_1^1,U_2^1,U_3^1, D_1^1,D_2^1,D_3^1, L_1^1, L_2^1, L_3^1, E_1^1, E_2^1, E_3^1, N_1^1, N_2^1,N_3^1\verb!},!\, 
\end{equation*}
\begin{equation}
\verb!{!Q_1^2,Q_2^2,Q_3^2, U_1^2,U_2^2,U_3^2, D_1^2,D_2^2,D_3^2, L_1^2, L_2^2, L_3^2, E_1^2, E_2^2, E_3^2, N_1^2, N_2^2,N_3^2\verb!}}]!
\end{equation}
As an example for the two sets of charges $\{BL,Y_3\}$, input
\begin{equation*}
\verb!testCharges[{{!1, 1, 1, -1, -1, -1, -1, -1, -1, -3, -3, -3, 3, 3, 3, 3, 3, 3\verb!},!\end{equation*}
\begin{equation}
\verb!{!0, 0, 1, 0, 0, -4, 0, 0, 2, 0, 0, -3, 0, 0, 6, 0, 0, 0\verb!}}]!
\end{equation}

An example output of the function \texttt{testCharges} is~\\~\\
\noindent \setlength{\tabcolsep}{4pt}
\begin{tabular}{|c|c|cccc|cc|cccc|}
\hline
Anomaly free& Vector-like & $P_2$ & $P_6$ & $P_7$ & $P_{16}$ & $P_{14}$ & $P_{22}$ & $P_{8}$ & $P_{15}$ & $P_{26}$ & $P_{27}$ \\
\hline
True &False& True & False & True & True & True& True & False & True & True& True \\\hline
\multicolumn{12}{|c|}{In a semi-simple extension free of local anomalies: True}\\\hline
\multicolumn{12}{|c|}{In a semi-simple extension free of local \& global anomalies: True}\\
\hline
\end{tabular}
~\\~\\
\noindent The information contained in this table is as follows:
\begin{itemize}
\item `Anomaly free': this returns `true' if and only if every point on the plane spanned by $\{X^n\}$ is anomaly free.
\item `Vector-like': this returns `true' if and only if every point in the set $\{X^n\}$ is vector-like (and therefore, every point on the plane spanned by $\{X^n\}$ is vector-like).
\item `$P_i$': this returns `true' if and only if the plane spanned by $\{X^n\}$ lies in the plane $P_i$. We have only specified this for the `maximal' planes in Eqs.~(\ref{eq:P2}--\ref{eq:P27}).
\item `In a semi-simple extension free of local anomalies': this returns `true' if and only if the plane spanned by $\{X^n\}$ sits in a semi-simple extension free of local anomalies \emph{i.e.} in (\ref{eq:FLLA}), and therefore the corresponding extension of the $\sm$ by $\u$'s sits in a semi-simple extension free of local anomalies. Put another way, it is true if and only if at least one of the $P_i$'s given returns true. 
\item `In a semi-simple extension free of local \& global anomalies': this is similar to the previous bullet, but now restricting attention to those semi-simple extensions that are furthermore free of global anomalies. We therefore require the plane spanned by $\{X^n\}$ to sit in (\ref{eq:FLGALA}). Thus, it returns true if and only if at least one of the $P_i$'s for $i=2,6,7,16, 8,14,26,27$ returns true ({\em i.e.} skipping the `middle' block of our output).
\end{itemize}

\subsection{Other programs}
We include several other programs in the ancillary directory to our \texttt{arXiv} submission. These are 
\begin{itemize}
\item \texttt{Test\_your\_own\_charges.cpp}: similar to \texttt{Test\_your\_own\_charges.nb}, this is a C\texttt{++} program which of the planes in Eqs.~(\ref{eq:P2}--\ref{eq:P27}) a single set of charges $X$ sits. From this it can be deduced weather it sits in any semi-simple extension absent of local anomalies or absent of local and global anomalies.
\item \texttt{Generate\_up\_to\_Qmax.cpp}: this is a C\texttt{++} program which given a maximum charge $\Qmax$ generates, and counts all charges in each of our 10 planes in Eqs.~(\ref{eq:P2}--\ref{eq:P27}) up to this $\Qmax$. This is related to our discussion in Section~\ref{sec:Atlas}.  
\item \texttt{Atlas\_scan.cpp}: this is a C\texttt{++} program that scans through the Anomaly Free Atlas of Ref.~\cite{Allanach:2018vjg}, and checks how many sit in each plane. Again, this is related to our discussion in Section~\ref{sec:Atlas}.
\item \texttt{Plane\_checks.h}: this is an auxiliary script for the C\texttt{++} programs above.
\item \texttt{Flatland.nb}: this is a \texttt{Mathematica} script in which much of the workings out in this paper have been explicitly written.
\item \texttt{Flatland\_aux.m}: this is a \texttt{Mathematica} script which is auxiliary to \texttt{Flatland.nb}. It contains functions related to the Chevalley basis, described in Appendix~\ref{app:chevalley}. 
\end{itemize}

\section{Summary }

In this paper, we introduced the notion of a general abelian extension of the Standard Model (including three right-handed neutrinos), and found all such abelian extensions that embed in anomaly-free semi-simple gauge theories, allowing for generic flavour non-universality. Such abelian extensions are widely used in model building, providing a gauge theory framework for extensions of the SM by heavy neutral $\ZP$ gauge bosons. 

This space of abelian extensions is a linear space, which we call `Flatland', composed of a set of planes whose dimensions do not exceed six. Consequently, starting from any such semi-simple gauge theory and breaking down to the SM, one can produce at most five neutral $\ZP$ bosons with linearly independent fermion charges (the offset with the dimensionality of Flatland is because hypercharge, which remains unbroken, is always included in the abelian part). Many different semi-simple gauge models can give this maximal number of independent $\ZP$ bosons; examples include $\su(12)\oplus \su(2)\oplus \su(2)$ and $\so(10)^{\oplus 3}$. We explicitly parametrise the planes that compose Flatland analytically. We prove various results along the way: for example, we show that every anomaly-free vector-like abelian extension of the SM necessarily embeds in an $\su(12)\oplus \su(2)\oplus \su(2)$ gauge model.

We provide a short computer program, described in Section~\ref{sec:program}, to enable users to test whether an abelian extension of their choosing (for example, their favourite anomaly-free $\ZP$ model) fits inside any semi-simple gauge model. The code moreover lists a set of maximal gauge algebras (if there are any) in which the user's model embeds. We applied this code to a selection of models from the literature, including all anomaly-free flavour non-universal $\sm \oplus \u_X$ extensions with maximum charge 10. We hope that this work provides a useful bridge for model builders wanting to connect $\ZP$ models, which have many uses, with unified gauge models that might describe physics in the UV.

\acknowledgments

We thank Ben Gripaios for discussions.
JD has received funding from the European Research Council (ERC) under the European Union’s Horizon 2020 research and innovation programme under grant agreement 833280 (FLAY), and by the Swiss National Science Foundation (SNF) under contract 200020-204428. JTS was supported in part by the U.S. National Science Foundation (NSF) grant PHY-2014071. 

\appendix
\section{The Chevalley basis} \label{app:chevalley}

In this Appendix we recall the definition of the Chevalley basis of a simple Lie algebra $\g$. The Chevalley basis of a semi-simple Lie algebra is then trivially obtained from the union of bases of its simple ideals. We also recognise that we are using the phrase `Chevalley basis' in a looser sense than is present elsewhere in the literature.

Let $\Delta(\g)$ be a set of simple roots of $\g$, given some choice of Cartan subalgebra, and let $\Phi(\g)$ be the corresponding set of roots. We further denote the weight lattice by $\Lambda(\g)$, which as a group is isomorphic to $\mathbb{Z}^n$, and the Killing form by $\kappa_\g$.  For any $\lambda \in \Lambda(\g)$ we can define $h_{\lambda}\in \mathfrak{h}_\g$ through the Killing form, {\em viz.} 
\begin{align}
\kappa_\g(h_{\lambda}, h^\prime)=\lambda^\vee(h^\prime)\, ,
\end{align}
where $\lambda^\vee=2\lambda/(\lambda,\lambda)$ is the co-root associated with $\lambda$, and $(\cdot, \cdot)$ is the standard inner product defined using the Killing form. It is easy to check that $\{h_{\lambda}\}_{\lambda\in \Delta(\g)}$ forms a basis of $\mathfrak{h}_\g$ which is dual to the fundamental weights. 

The Lie algebra $\g$ has the basis $\{h_{\lambda}\}_{\lambda\in \Delta(\g)}$ and $\{e_{\tilde \lambda}\}_{\lambda\in \Phi(\g)}$. To specify the commutators in this basis, it is convenient to split our discussion into two cases; that of simply laced and not simply laced simple Lie algebras.

\paragraph{Simply laced:} If $\g$ is simply laced algebras, meaning all the roots are of the same length, then the commutators are~\cite{Frenkel:1980rn,kac1998vertex,frenkel2004vertex}, for all $\lambda_1,\lambda_2\in \Delta(\g)$ and $\tilde \lambda, \tilde \lambda_1,\tilde \lambda_2 \in \Phi(\g)$,
\begin{align}
[h_{\lambda_1},h_{\lambda_2}]&=0\, ,\label{eq:chevalley1}\\
[h_{\lambda_1},e_{\tilde \lambda}]&=(\lambda_1,\tilde \lambda) e_{\tilde \lambda}\, ,\\
[e_{\tilde \lambda}, e_{-\tilde \lambda}]&=\epsilon(\tilde \lambda, -\tilde \lambda)h_{\tilde \lambda}\, ,\\
[e_{\tilde \lambda_1},e_{\tilde \lambda_2}]&=\begin{cases}\epsilon(\tilde \lambda_1,\tilde \lambda_2)e_{\tilde \lambda_1+\tilde \lambda_2}& \text{if $\tilde \lambda_1+\tilde \lambda_2\in \Phi(\mathfrak{g})$\, ,}\\
0& \text{otherwise .}\end{cases} \label{eq:chevalley4}
\end{align}
Here $\epsilon:\Lambda(\mathfrak{g})\times \Lambda(\mathfrak{g})\rightarrow \mathbb{C}^\ast$ is any cocycle in the group cohomology class of $H^2(\Lambda(\mathfrak{g}),\mathbb{C}^\ast)$ uniquely defined by the condition that~\cite{frenkel2004vertex}
\begin{align}
\epsilon(\alpha,\beta)\epsilon(\beta,\alpha)&=e^{\pi i \left(\alpha,\beta\right)}\, ,
\end{align}
which must moreover satisfy the cocycle condition:
\begin{align}
\epsilon(\alpha+\beta,\gamma)\epsilon(\alpha,\beta)=\epsilon(\alpha,\beta+\gamma)\epsilon(\beta,\gamma),\quad \epsilon(1,1)=1\, .
\end{align}
Any such cocycle would work to define a basis. However, it is convention to choose one with $\mathrm{Im}\epsilon\in \{\pm1\}$, and such that
$\epsilon(\alpha,-\alpha)=-1$, $\epsilon(\alpha,\beta)=\epsilon(-\alpha,-\beta)$.
In this case the map $h_{\lambda}\mapsto -h_{\lambda}$, $e_{ \lambda}\mapsto e_{- \lambda}$ is an isomorphism of Lie algebras. In what follows, we shall not impose these conditions.

One can choose a different Chevalley basis for the Lie algebra $\g$, which corresponds to a different choice of cocycle $\epsilon^\prime$. The two bases, defined by cocycles $\epsilon$ and $\epsilon^\prime$, are then related as follows.
There will be a 1-cochain $c:\Lambda(\g)\rightarrow \mathbb{C}^\ast$ such that $\delta c \cdot\epsilon= \epsilon^\prime$, where the coboundary $\delta c: \Lambda(\g)\times \Lambda(\g)\rightarrow \mathbb{C}^\ast$ is defined through
\begin{align}
\delta c(\lambda_1,\lambda_2)=\frac{c(\lambda_1)c(\lambda_2)}{c(\lambda_1+\lambda_2)}.
\end{align} 
The change of basis then corresponds to 
\begin{align}
h_{\lambda} \mapsto h_{\lambda}, \quad e_{\lambda}\mapsto c(\lambda)e_{\lambda}.
\end{align}
\paragraph{Not simply laced:} Let us now turn our attention to the case of algebras which are not simply laced, an expos\'e of which can be found in~\cite{mitzman1985integral}. We focus entirely on the particular class of non-simply laced algebras $\mathfrak{sp}(2n)$, since no algebras of the form $\mathfrak{so}(2n+1)$ appear in this paper.

The trick here is to use the simply laced algebra $\mathfrak{su}(2n)$ as a starting point, and form $\sp(2n)$ as a subalgebra of it. There exists a (unique) non-trivial automorphism of weight lattices $\sigma:\Lambda(\mathfrak{su}(2n))\rightarrow \Lambda(\mathfrak{su}(2n))$, which is induced by the non-trivial graph automorphism of the Dynkin digram of $\mathfrak{su}(2n)$. Using this automorphism, we can define the subsets
\begin{align}
\Delta_{[0]}=\left\{\frac{1}{2}(\lambda+\sigma(\lambda))\mid \lambda \in \Delta(\mathfrak{g})\right\}, \quad \Phi_{[0]}=\left\{\frac{1}{2}(\lambda+\sigma(\lambda))\mid \lambda \in \Phi(\mathfrak{g})\right\}.
\end{align}
It is not hard to show that $\Phi_{[0]}$ is isomorphic to $\Phi(\sp(2n))$ and that $\Delta_{[0]}$ is a valid set of simple roots of $\Phi_{[0]}$. We explicitly denote this isomorphism by $\tau:\Phi(\sp(2n))\rightarrow \Phi_{[0]}$. 
From this we can use $\{h^\prime_{\lambda}\}_{\lambda \in \Delta(\sp(2n))}$ as a Cartan subalgebra of $\sp(2n)$, where
\begin{align}
h^\prime_{\lambda}=h_{\tau(\lambda)^\vee}\, .
\end{align}

We now want to complete our construction of the basis by determining what form $\{e^\prime_{\lambda}\}_{\lambda\in \Phi(\sp(2n))}$ should take. This depends on our choice of cocycle $\epsilon$, and will also depend on a few other choices. In particular, there always exists a 1-cochain $c$ such that 
\begin{align}
\delta c(\sigma(\lambda_1),\sigma(\lambda_2)) \epsilon(\sigma(\lambda_1),\sigma(\lambda_2))=\delta c(\lambda_1,\lambda_2) \epsilon(\lambda_1,\lambda_2)\, .
\end{align}
This condition does not fix the 2-cocycle $\delta c$, and so this freedom allows different $\delta c\cdot \epsilon$.
Different choices of the 2-cocycle $\epsilon$ and the 1-cocycle $c$, satisfying the conditions we have stated, will nevertheless lead to isomorphic subalgebras. We then have that, for $\lambda\in \Phi(\sp(2n))$, 
\begin{align}
e^\prime_{\lambda}=\begin{cases}c(\tilde \lambda) e_{\tilde\lambda}+c(\sigma(\tilde \lambda))e_{\sigma(\tilde\lambda)} & \text{if $\sigma(\tilde\lambda)\ne \tilde\lambda$}\, ,\\
c(\tilde \lambda)e_{\tilde\lambda} & \text{if $\sigma(\tilde\lambda)=\tilde\lambda$}\, ,\end{cases}
\end{align}
where $\tilde \lambda$ is either element of $\Phi(\su(2n))$ such that $\tau(\lambda)=\frac{1}{2}(\tilde \lambda+\sigma(\tilde \lambda))$.

The fact that $\{h^\prime_{\lambda}\}_{\lambda \in \Delta(\sp(2n))}$ and $\{e^\prime_{\lambda}\}_{\lambda\in \Phi(\sp(2n))}$ then span a subalgebra of $\su(2n)$ isomorphic to $\sp(2n)$ is proven in~\cite[Thm. 3.2.6]{mitzman1985integral}, and we refer the interested reader to this reference for more details.

\subsection{Examples: $\su(2)$, $\su(3)$ and $\su(4)$} \label{app:chevalley_su4}
\paragraph{The algebra $\su(2)$:} As mentioned in the main text, the root system of $\su(2)$ is just $\Phi(\su(2))=\{\pm \kappa\}$, with the simple root $\Delta(\su(2))=\{\kappa\}$. To specify the commutators in Eqs.~(\ref{eq:chevalley1}--\ref{eq:chevalley4}) we need to do two things. Firstly, we specify the inner product $(\cdot, \cdot)$, which is simply
\begin{align}
(\kappa,\kappa)=2.
\end{align}  
Secondly, we must make a choice of 2-cocycle  $\epsilon_{\su(2)}: \mathbb{Z}\times \mathbb{Z}\rightarrow \mathbb{C}^\ast$. We choose
\begin{align}
\epsilon_{\su(2)}(a_1 \kappa, b_1 \kappa)=(-1)^{a_1 b_1}\, , \qquad (a_1, b_1) \in \Z^2\, .
\end{align}
{Explicitly, this gives the $\su(2)$ commutation relations: 
\begin{align}
[h_{\kappa}, e_{\pm \kappa}]=2 e_{\pm \kappa}\,, \quad [ e_{-\kappa},e_{\kappa}]=h_{\kappa}\,.
\end{align}

\paragraph{The algebra $\su(3)$:} The simple roots of $\su(3)$ are $\Delta(\su(3))=\{\tau_1,\tau_2\}$, whilst the full root system is 
\begin{align}
\Phi(\su(3))=\{ \pm\tau_1,\pm\tau_2,\pm(\tau_1+\tau_2)\}.
\end{align}
As is standard, the inner product is given by 
\begin{align}
(\tau_1,\tau_1)=2,\quad (\tau_2,\tau_2)=2,\quad (\tau_1, \tau_2)=-1.
\end{align}
We make the choice of 2-cocycle $\epsilon_{\su(3)}: \mathbb{Z}^2\times \mathbb{Z}^2\rightarrow \mathbb{C}^\ast$ to be 
\begin{align}
\epsilon_{\su(3)}(a_1 \tau_1+a_2\tau_2,b_1\tau_1+b_2\tau_2)=(-1)^{b_1(a_1-a_2)} (-1)^{a_2 b_2}.
\end{align}
{We will not write out the commutators here explicitly, since the number of equations are long. They can easily be deduced from the above, and Eqs.~(\ref{eq:chevalley1}--\ref{eq:chevalley4}).}

\paragraph{The algebra $\su(4)$:} For $\su(4)$, let us use the notation $\Delta(\su(4))=\{\mu_1,\mu_2,\mu_3\}$ for the simple roots. (This differs from the notation used in the main text in Subsections~\ref{sec:so6I} and~\ref{sec:so6II}; however, the present notation is convenient for the next Subsection of this Appendix.) The root system is then 
\begin{align}
\Phi(\su(4))=\{ \pm \mu_1,\pm \mu_2,\pm \mu_3, \pm(\mu_1+\mu_2),\pm (\mu_2+\mu_3),\pm(\mu_1+\mu_2+\mu_3)\}.
\end{align}
The inner products are 
\begin{align}
(\mu_1,\mu_1)=2, \quad (\mu_2,\mu_2)=2,\quad (\mu_3,\mu_3)=2,\nonumber\\
(\mu_1,\mu_2)=-1,\quad (\mu_2,\mu_3)=-1,\quad (\mu_1,\mu_3)=0.
\end{align}
For the 2-cocycle $\epsilon_{\su(4)}:\mathbb{Z}^3\times \mathbb{Z}^3\rightarrow \mathbb{C}^\ast$ we choose 
\begin{align}
\epsilon_{\su(4)}(c_1\mu_1+c_2\mu_2+c_3\mu_3,d_1\mu_1+d_2\mu_2+d_3\mu_3)=(-1)^{d_1(c_1-c_2)}(-1)^{d_2 c_2} (-1)^{d_3(c_3-c_2)}.
\end{align} 

\subsection{Example: $\mathfrak{sp}(4)$}
For the non simply laced example of $\mathfrak{sp}(4)$, we transfer our results from $\mathfrak{su}(4)$. The automorphism $\sigma: \Lambda(\su(4))\rightarrow \Lambda(\su(4))$ is
\begin{align}
\sigma(c_1\mu_1+c_2\mu_2+c_3\mu_3)=c_1\mu_3+c_2\mu_2+c_3\mu_1.
\end{align}
The simple roots are then given by $\Delta_{[0]}=\{\frac{1}{2}(\mu_1+\mu_2),\mu_2\}$. Denoting the simple roots of $\sp(4)$ as $\Delta(\sp(4))=\{\lambda_1, \lambda_2\}$, the automorphism $\tau:\Phi(\sp(4))\rightarrow \Phi_{[0]}$ restricted to $\Delta(\sp(4))$ takes the form 
\begin{align}
\tau(\lambda_1)=\frac{1}{2}(\mu_1+\mu_2)\,,\qquad \tau(\lambda_2)=\mu_2\,.
\end{align}
From this alone we can see that  
\begin{align}
{h}_{{\lambda}_1}^\prime&:=h_{\mu_1}+h_{\mu_3}\, ,\\
{h}_{\lambda_2}^\prime&:=h_{\mu_2}\, .
\end{align}
The full root system of $\sp(4)$ is
\begin{align}
\Phi(\sp(4))=\{ \pm{\lambda}_1,\pm{\lambda}_2,\pm(2{\lambda}_1+{\lambda}_2),\pm({\lambda}_1+{\lambda}_2)\}\,,
\end{align}
and the form of $\Phi_{[0]}$ and $\tau$ on $\Phi(\sp(4))$ can be deduced by linearity. 

We took care to choose a 2-cocycle $\epsilon_{\su(4)}$ above that is invariant under pull-back along the automorphism $\sigma$.
Thus, when choosing a 1-cochain $c$ to define $e_\lambda^\prime$, we can simply choose the constant map to $1\in \mathbb{C}^\ast$. This means we have 
\begin{align}
{e}_{\pm{\lambda}_1}^\prime&:=e_{\pm\mu_1}+e_{\pm\mu_3}\,, \nonumber\\
{e}_{\pm{\lambda}_2}^\prime&:=e_{\pm\mu_2}\,,\nonumber \\
{e}_{\pm({\lambda}_1+{\lambda}_2)}^\prime&:=e_{\pm\mu_{12}}+e_{\pm\mu_{23}}\,, \nonumber \\ 
{e}_{\pm(2{\lambda}_1+{\lambda}_2)}^\prime&:=e_{\pm\mu_{123}}\, ,
\end{align}
where {\em e.g.} $\mu_{123}=\mu_1+\mu_2+\mu_3$.
Thus $\{h_{\lambda_1}^\prime,h_{\lambda_2}^\prime\}\cup \{{e}_{\pm{\lambda}_1}^\prime,{e}_{\pm{\lambda}_2}^\prime,{e}_{\pm({\lambda}_1+{\lambda}_2)}^\prime,{e}_{\pm(2{\lambda}_1+{\lambda}_2)}^\prime\}$ forms a basis of $\sp(4)$. In the main text we drop the primes when denoting the elements of this basis.

\bibliographystyle{JHEP} 
\bibliography{references}
\end{document}